\documentclass[useAMS,usenatbib]{mn2e}
\usepackage{amsmath}
\usepackage{amssymb}
\usepackage[pdftex]{graphicx}
\title{Investigating the effects of chemistry on molecular line profiles  of infalling low mass cores}
\author[J. F. Roberts, J. M. C. Rawlings and H. A. Stace]{J. F. Roberts$^{1,2}$\thanks{E-mail: robertsj@inta.es}, J. M. C. Rawlings$^2$ and H. A. Stace$^2$\\
$^1$ Centro de Astrobiolog\'ia (CSIC/INTA), Ctra de Torrej\'on a Ajalvir km 4, E-28850 Torrej\'on de Ardoz, Madrid, Spain\\
$^2$ Department of Physics \& Astronomy, University College London, Gower Street, London WC1E 6BT, UK}

\begin{document}
\date{July 2010}
\pagerange{\pageref{firstpage}--\pageref{lastpage}} 

\maketitle
\label{firstpage}

\begin{abstract}
We have coupled a chemical model with two dynamical models of collapsing low mass star-forming cores to predict abundances across the core of the commonly used infall tracers, CS and HCO$^+$, at various stages of the collapse. The models investigated are a new ambipolar diffusion model and the `inside-out'  collapse model. 
We have then used these results as an input to a radiative transfer model to predict the line profiles of several transitions of these molecules. 
 For the inside-out collapse model, we predict significant molecular depletion due to freeze-out in the core centre, which prevents the formation of the blue asymmetry (believed to be the `signature' of infall) in the line profiles. 
 Molecular depletion also occurs in the ambipolar diffusion model during the late stages of collapse,  but the line profiles still exhibit a strong blue asymmetry due to extended infall.
For the inside-out collapse model to exhibit the blue asymmetry it is necessary to impose a negative kinetic temperature gradient on the core and suppress freeze-out.
 Since freeze-out is observed in several class 0 protostars which are thought to be collapsing, this presents a major inconsistency in the inside-out collapse model of star formation.
\end{abstract}

\begin{keywords}
astrochemistry -- line:profiles -- stars:formation -- ISM:molecules
\end{keywords}

\section{Introduction}






 
The current method for detecting infall in pre-stellar cores and young stellar objects (YSOs) involves observing an optically thick  line, such as CS J=2$\to$1 and 3$\to$2, along with an optically thin  line, such as N$_2$H$^+$ J=1$\to$0 (both of which are tracers of dense gas) towards the selected core (Lee et al. 2004).  If the optically thick lines exhibit a double peaked profile, with the blue-shifted peak stronger than the red-shifted peak (i.e. a {\it blue asymmetry}), and the optically thin line is single peaked  with a central velocity that lies inbetween the two peaks of the optically thick transition (to verify that the double peak does not arise from two separate dense clouds along the line of sight), then it is a strong, but not unambiguous, indicator that the core is collapsing.  

The blue asymmetry is not unique to a particular collapse model (Leung \& Brown 1977), although many astronomers interpret their observations of both spectral line and continuum emission by comparison with the inside-out collapse model of Shu (1977) (e.g. Zhou et al 1993; Motte \& Andre 2001).  The inside-out collapse model is also widely used in theoretical studies of collapsing cores (e.g. Tsamis et al. 2008; Rawlings \& Yates 2001; Rawlings et al. 1992; Evans et al. 1994; Weidenschilling et al. 1994).

However, there are several theoretical arguments and observational studies which indicate that the inside-out collapse model may be inappropriate.  For example, 
 \citet{Belloche02} found that the class 0 protostar IRAM 04191
shows evidence of extended inward motions, which are inconsistent with the model of Shu (1977), which predicts that the outer parts of the core should be static.

The inside-out collapse model of \citet{Shu77} describes the protostellar phase, and begins from a singular isothermal sphere.  Preceding this is the pre-protostellar phase (often referred to as the pre-stellar phase for breivity; \cite{WT94}).  It has been suggested that during this time a diffuse core evolves quasi-statically into a more centrally condensed pre-stellar core via the process of ambipolar diffusion.
In this paper, we  model the line profiles of various transitions of the commonly used infall tracers CS and HCO$^+$ produced by two different models of collapsing cores:
(i) a new ambipolar diffusion model (Stace \& Rawlings 2010, in preparation; hereafter SR10) and (ii)  the standard inside-out collapse model (Shu 1977).  
 The ambipolar diffusion model covers the pre-stellar phase, but by $\sim10^6$~yr it 
extends into conditions similar to the protostellar regime, reaching a density of $n_\textrm{H}\sim2\times10^7$~cm$^{-3}$ at a distance of $10^{16}$~cm ($\sim700$AU) from the core centre, which is actually higher than that estimated for the protostellar core  IRAM 04191 at such a distance \citep{Belloche02}
At this time, 
 one of the main differences between these two models is that the inside-out collapse model has a static envelope, whereas in the ambipolar diffusion model there are extended low-velocity inward motions.

 An important aspect of this work is that we couple the dynamical models to a large chemical network, in order to predict more accurate abundance profiles of each molecule.  These abundance profiles are then used as an input to a radiative transfer model, along with the density and velocity profiles appropriate for each collapse model.  
 We focus on how the chemical evolution of the cores affects the resulting line profiles shapes, and we find that the chemical distribution can be an important factor in determining whether or not the blue-asymmetry is seen.

 The paper is organised as follows.  In Section~\ref{models} we describe dynamical collapse models studied in the paper, and in Section~\ref{sec:chemmod} we describe the chemical model, the results of which  are given in Section~\ref{chemdisc}.   In Section~\ref{GLP} we describe the radiative transfer model used and how the line profiles were generated, and in Section~\ref{RAD} we present the resulting line profiles, and also discuss their sensitivity to the temperature profile of the cores.  In Section~\ref{collapse_disc} we present a discussion of the results, considering their sensitivity to the conditions chosen in the chemical model.  The conclusions are summarised in Section~\ref{conc}.

\section{Dynamical models}
\label{models}
\subsection{Ambipolar diffusion}
\label{ADcoll}

Ambipolar diffusion controlled collapse occurs in magnetically subcritical cores, where the magnetic field is strong enough to support the core against gravitational collapse.  
It is estimated that a core is subcritical if the core mass to magnetic flux ratio, $M/\phi$, is less than $\sim 0.13G^{-1/2}$  (in cgs units), where $G$ is the gravitational constant (Mouschovias \& Spitzer 1976).
In such cores, the neutral species do not directly feel the effect of the magnetic field, and therefore drift towards the centre of core, being impeded by the drag force exerted by the ions which are tied to the magnetic field.  Eventually, in the centre of the core, enough mass builds up so that the core becomes supercritical and begins to collapse on a dynamical timescale.
 
We have taken preliminary results from the dynamical model from SR10 to couple with our chemical model.    
The model will be described in detail in SR10, but the following is a brief summary:

The dynamical model follows the spherically symmetric collapse of a core composed of 100 concentric shells of gas and dust.  The core is embedded in a cylindrical magnetic field, which evolves throughout the collapse, although its geometry always remains cylindrical.  There is no mass transfer between the shells.

At each time step, the code calculates the total force exerted on each shell of gas, taking into account the gravitational force, the gas pressure and the drag force exerted on the neutral species by the ions and charged dust grains.  These drag forces depend on the number density of charged particles, and on the drift velocity between ions and neutrals, which is related to the magnetic field strength.  To calculate the magnetic field at each step, it is assumed that the magnetic field is frozen into the plasma.

The code also calculates the visual extinction for each shell, which increases during the collapse due to the increase in the density of dust grains.  A relationship between visual extinction and temperature is assumed, so the radial temperature profile also evolves throughout the collapse, which affects the gas pressure.  The temperatures in the core ranges from 6.5~K to 13.5~K, although 
 since in our chemical model it is not yet possible to include a time dependent chemistry, 
we have assumed a constant gas temperature of 10~K.
 The chemical abundances are not significantly affected by this simplification; running the chemical model at temperatures of 6.5~K and 13.5~K results in differences in the CS and HCO$^+$ abundances of $\lesssim30$\%.


In the model of SR10, account is taken of how the changing chemical composition of the core affects the collapse dynamics;  chemical reactions and freeze-out of ions onto dust grains affect the overall ionisation fraction of the gas, which therefore alters the drag force experienced by the neutrals.  The rates of the chemical reactions depend on the density, visual extinction and temperature calculated by the dynamical equations.  SR10 have coupled the chemical and dynamical equations to build a self-consistent model of ambipolar diffusion controlled collapse.

We have used the results for a core of mass 6.80~$M_\odot$, radius $8.45 \times 10^{17}$~cm and magnetic field strength  4 $\mu$G,  leading to a mass to flux ratio of $M/(B\pi R^2)\sim1500$~g~G$^{-1}$~cm$^{-2}$.
This is $\sim 3$ times the critical value calculated by \citet{Mousc76}, consistent with observations of molecular clouds \citep{Crutcher99}.  Although this is a `supercritical' core according to the \citet{Mousc76} criterium, the magnetic field is still strong enough to significantly retard the collapse.
The initial density distribution of the core is uniform, with $n_\text{H}=2.8 \times 10^3$~cm$^{-3}$.
The  data we have taken from this model are: (i) radius as a function of time, and (ii) density and velocity as a function of time and radius, for the set of 100 infalling shells making up a collapsing core.  We run chemical models for six of these shells, spanning the entire core radius.  
We derive the chemical evolution of the shells 
with 
starting radii of $8.45 \times 10^{17}$~cm, $7.37 \times 10^{17}$~cm, $6.77 \times 10^{17}$~cm, $ 5.41 \times 10^{17}$~cm, $ 3.93 \times 10^{17}$~cm and $1.82 \times 10^{17}$~cm respectively.

\begin{table}
\caption{Table showing the initial radius ($R_0$) and density ($n_{\text{H}0}$)  for each collapsing shell for which we derive the chemical evolution, for the ambipolar diffusion and inside-out collapse models. }
\begin{center}
\begin{tabular}{cc|cc}
\hline \hline
\multicolumn{2}{c|} {{\sc Ambipolar Diffusion}} & \multicolumn{2}{c} {{\sc Inside-out}} \\ \hline \hline
\multicolumn{2}{c|} {Total mass 6.80~$M_\odot$ } & \multicolumn{2}{c} {Total mass 0.96~$M_\odot$} \\ 
\multicolumn{2}{c|} {Outer radius $8.45 \times 10^{17}$~cm } & \multicolumn{2}{c} {Outer radius $1.60 \times 10^{17}$~cm} \\ \hline \hline
$R_0$ (cm) & $n_{\text{H}0}$ (cm$^{-3}$)  & $R_0$ (cm) & $n_{\text{H}0}$ (cm$^{-3}$) \\ \hline
$1.82 \times 10^{17}$ & $2.8 \times 10^3$  & $1.0 \times 10^{16}$ & $4.7 \times 10^{6}$ \\
$3.93 \times 10^{17}$ & $2.8 \times 10^3$  & $2.5 \times 10^{16}$ & $7.5 \times 10^{5}$ \\
$5.41 \times 10^{17}$ & $2.8 \times 10^3$  & $5.0 \times 10^{16}$ & $1.9 \times 10^{5}$ \\
$6.77 \times 10^{17}$ & $2.8 \times 10^3$  & $7.4 \times 10^{16}$ & $8.7 \times 10^{4}$ \\
$7.73 \times 10^{17}$ & $2.8 \times 10^3$ & $1.6 \times 10^{17}$ & $1.8 \times 10^{4}$  \\
$8.45 \times 10^{17}$ & $2.8 \times 10^3$  &   \multicolumn{2}{c} {}\\
\end{tabular}
\end{center}
\label{IPCs}
\end{table}

\begin{figure}
\includegraphics[width=90mm]{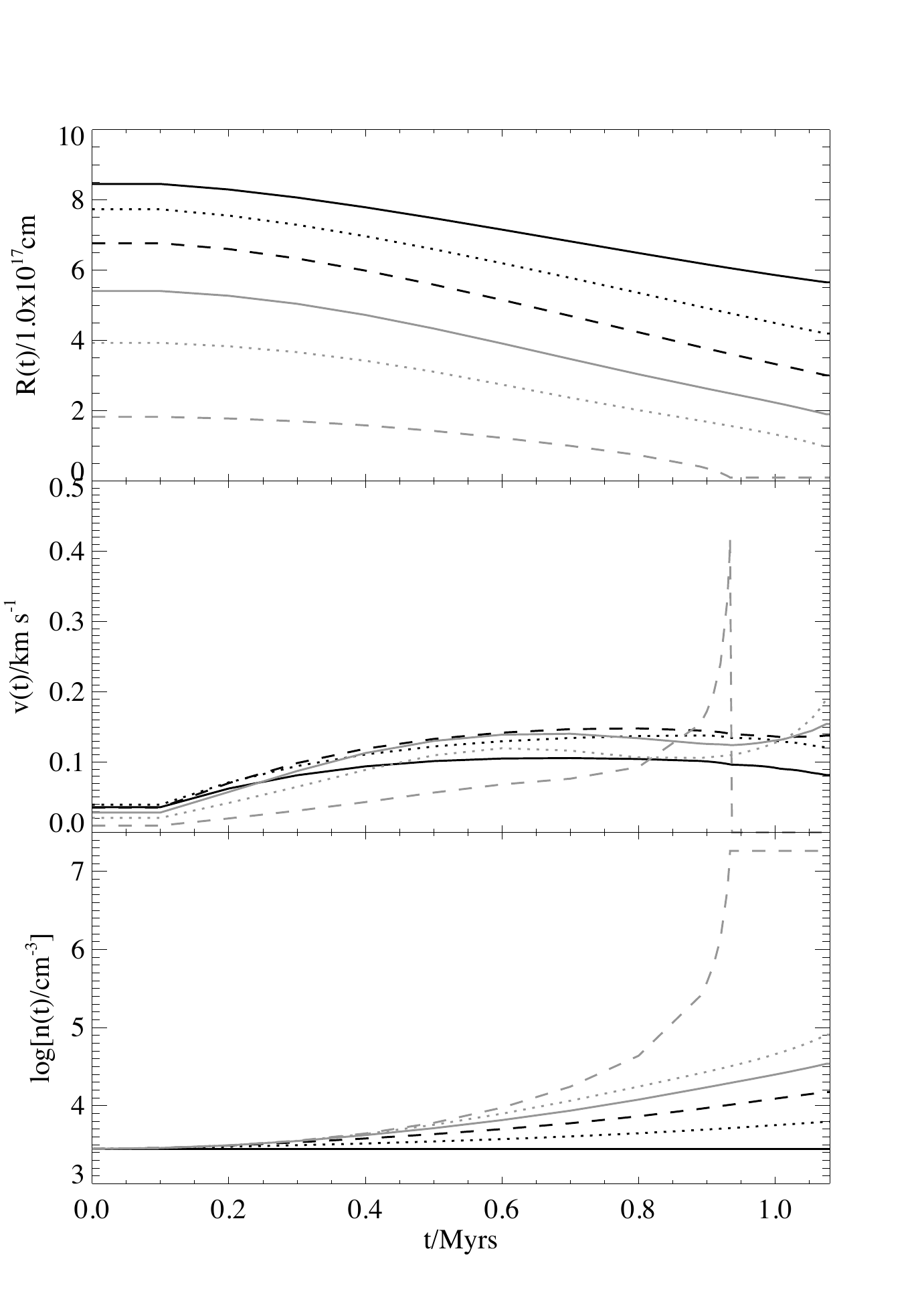}
\caption{Physical evolution (radial position, velocity and density) of the test shells in the ambipolar diffusion model.  The black solid line, black dotted line, black dashed line, grey solid line, grey dotted line and grey dashed line are for the shells beginning at $R_0 = 8.45 \times 10^{17}$, $7.73 \times 10^{17}$, $6.77 \times 10^{17}$, $5.41 \times 10^{17}$, $3.93 \times 10^{17}$ and $1.82 \times 10^{17}$~cm respectively.}
\label{fig:ambdiff_phys}
\end{figure}

The physical evolution of the test shells is shown in Figure~\ref{fig:ambdiff_phys}.  
 The core contracts with time (see the evolution of the outer shell, plotted with the black solid line in Figure~\ref{fig:ambdiff_phys}), as the density of the inner shells increases.
The inner shell (grey dashed line),  collapses relatively quickly into the core, and before $10^6$~yr it reaches its final density of $\sim 10^7$~cm$^{-3}$  where the collapse halts  for this shell.
 This is because for such high and rapidly changing densities, the  model of SR10 cannot converge on a solution. 

Since the dynamics of SR10's model depend upon the chemistry, particularly the ionisation fraction of the gas, it is possible that a different set of chemical reactions and initial conditions could be inconsistent with the dynamics predicted by SR10's models.  
The set of reactions used in the chemical model of SR10 was taken from the UMIST RATE95 database (Millar et al. 1997), whereas in our model we have used a more up-to-date reaction set from the RATE06 database (Woodall et al. 2007) (see Section~\ref{sec:chemmod}). 
 Furthermore, SR10 calculated the visual extinction self consistently with grain distribution in the core, whereas for simplicity we assumed a constant visual extinction, which will have an effect on the rates of photoreactions, particularly at the edge of the core.

Our chemical model predicts an ionisation fraction of $\sim 10^{-6}$ throughout the core at early times ($\sim 10^5$~yrs), which is approximately an order of magnitude higher than the ionisation fraction predicted by the SR10 model.  This means that for the earlier stages of the collapse, the gas in our model should be more strongly coupled to the magnetic field, which would mean there is more resistance to the collapse and the length of time spent at lower densities would increase,
 effectively introducing an almost static phase before the initiation of the collapse. 
 In Section~\ref{abundances} we investigate the effect of including such a static phase, and we find that it only has an effect on the resulting line profiles if the static phase exists for $> 10^{5}$~yr.
%
%
 Once the collapse is underway, the
difference in ionisation fractions between the two models reduces: At $5\times10^5$~yr, 
our model predicts ionisation fractions of approximately 2.5$\times$ larger than those
of SR10, and at $1\times10^6$~yr, for radii $\leq4\times10^{17}$~cm, the ionisation fractions of
our model are a factor of $1.3-2$$\times$ smaller than those of SR10.  For radii $>4\times10^{17}$~cm,
the ionisation fractions of our model are factor of $2-4\times$ smaller than those of SR10. 
Since a factor of 4 difference in the ionisation fraction might be significant,
as it means the gas in our model would be less strongly coupled to the 
magnetic field compared to the gas in SR10's model and may therefore have a higher 
infall velocity, in Section~\ref{IF} we test what happens to the CS line profiles when the velocity of the gas at large radii ($>4\times10^{17}$~cm) is increased.

 Note that we find that our chemical model predicts ionisation fractions consistent with observations for late times: The estimated ionisation fractions (with respect  to H nuclei) of the pre-stellar cores B68 and L1544 are $\sim5\times10^{-9}$  (for the central regions where $A_V\gtrsim5$) and $\sim5\times10^{-10}$  (for radii $\leq2500$AU) respectively \citep{Maret07,Caselli02}, which agree reasonably well with our  estimate of $\sim2\times10^{-9}$ in the core centre  ($\leq 1000$AU).



\subsection{Inside-out (collapse expansion wave) collapse}
\label{CEWdescription}
  Several solutions have been derived for the collapse of a self-gravitating isothermal sphere (e.g. \citet{Larson69,Penston69,Hunter77,Whitworth85}).  
  Here we concentrate on the commonly used solution derived by Shu (1977), which was solved by
%
finding similarity solutions to the continuity (mass conservation) equation and force equation for an ideal isothermal flow.
The collapse expansion wave solution is a particular solution to this problem, which starts with the initial density distribution of a singular isothermal sphere and with an initial velocity of zero everywhere.
This initial configuration is in unstable equilibrium, and so
a small perturbation at $t=0$~yr in the core centre can cause the central regions to begin collapsing.  The regions outside the collapsing centre have an unchanged density and gravitational field, so they can maintain hydrostatic equilibrium and are unaffected.  Material immediately above the collapsing inner region begins to fall, once the material beneath it has given way.  In this way, a `collapse expansion wave' (CEW) is formed, where material inside the CEW is falling, and material outside is static.  The CEW moves at the isothermal sound speed, $a$, which is the maximum speed that the collapsing inner regions can communicate with the static outer regions.  The position of the head of the CEW is therefore given by $r=at$ at time $t$.  The tail of the CEW is at the centre, where the material is approaching free-fall.  
This solution is valid for a singular isothermal sphere extending to infinity.  For a bounded sphere, the solution breaks down once the CEW hits the boundary.  

\begin{figure}
\includegraphics[width=90mm]{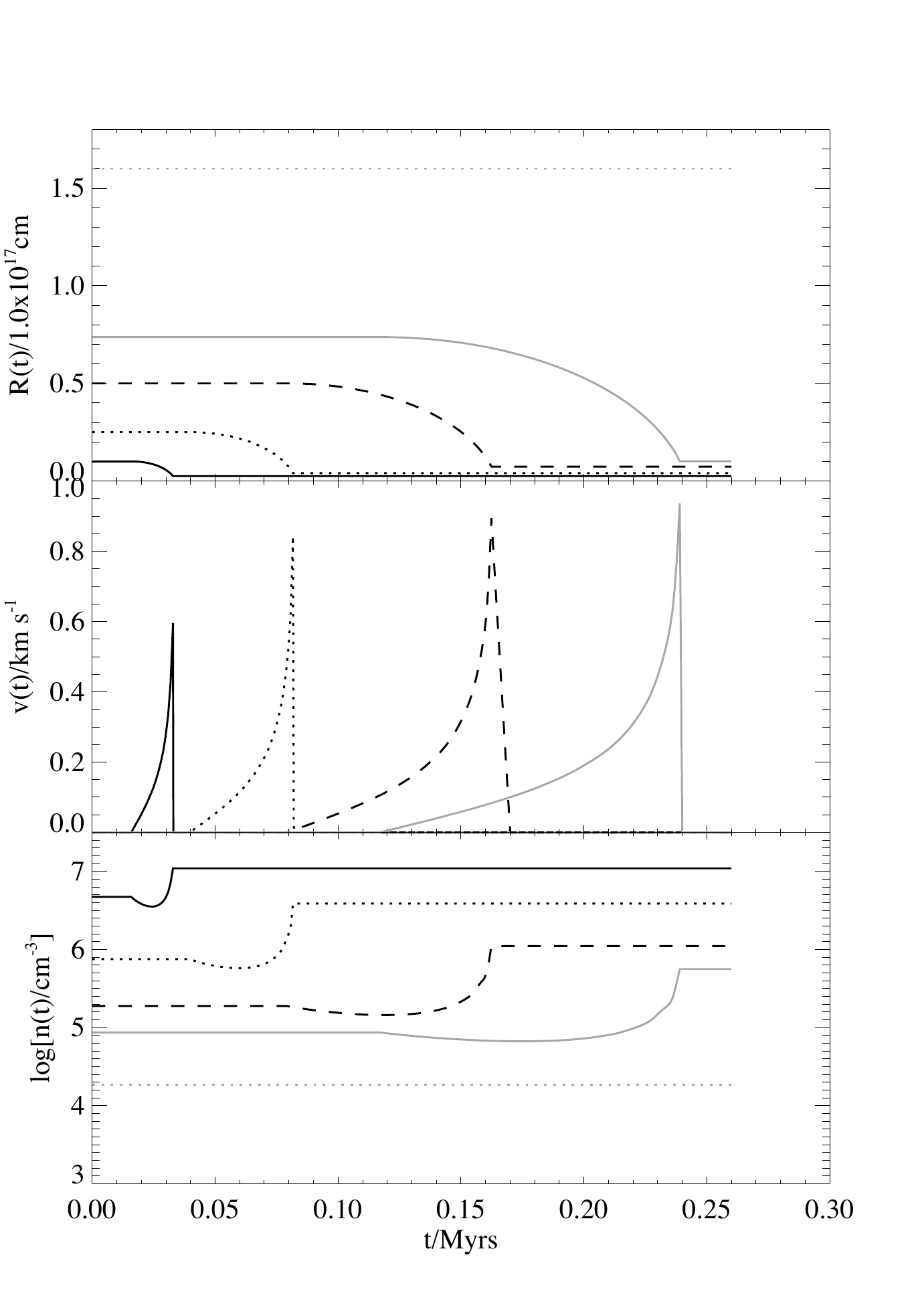}
\caption{Physical evolution (radial position, velocity and density) of the test shells in the inside-out collapse model.  The grey dotted line, grey solid line, black dashed line, black dotted line and black solid line refer to the  shells which start at $R_0 = 1.6 \times 10^{17}$, $7.4 \times 10^{16}$, $5.0 \times 10^{16}$, $2.5 \times 10^{16}$ and $1.0 \times 10^{16}$~cm respectively. }
\label{fig:cew_phys}
\end{figure}

As in the numerical example given by Shu (1977), we have used a core mass of 0.96~$M_\odot$, a core radius $1.6 \times 10^{17}$~cm and an isothermal sound speed of 0.2~km s$^{-1}$. 
 These values are applicable to Bok globules which are possible candidates for star formation.  The collapse finishes at $t=2.5 \times 10^5$~yr, when the CEW hits the boundary of the sphere.  The gas temperature of the core, used for the chemical calculations, is 10~K.
We ran the chemical model for a range of `test shells', starting at radii of $1.0 \times 10^{16}$~cm, $2.5 \times 10^{16}$~cm, $5.0 \times 10^{16}$~cm, $7.4 \times 10^{16}$~cm and $1.6 \times 10^{17}$~cm. 
In each case we calculated the density, velocity and position of the shell as functions of time using a spline interpolation of the CEW solution given in Table~2 of Shu (1977). 
Once  a shell gets close to the core centre, the density and velocity rapidly tend towards infinity and the solution is no longer valid.  At this point, we stop the collapse and set the velocity to zero.  
The radius, density and velocity of each shell as a function of time is shown in Figure~\ref{fig:cew_phys}.   The point when the collapse of each shell halts 
is clearly seen by the sudden drop to zero in velocity.  This point occurs once each shell reaches a distance of less than $\sim 1 \times 10^{16}$~cm from the centre of the core.  
The outer shell, at $1.6 \times 10^{17}$~cm remains static and at a constant density throughout the collapse.  This is because the solution ceases to be valid once the CEW has reached the outermost shell, so we stop the entire collapse at this point, which in our model occurs at $\sim 2.5 \times 10^5$~yr.
Table~\ref{IPCs}  shows the initial radius and density for each test shell for which we derive the chemical evolution, for both the ambipolar diffusion and inside-out collapse models.

 Note that the core mass and radius for the ambipolar diffusion model ($M=6.80$~$M_\odot$ and $R=8.45 \times 10^{17}$~cm) are much larger than that of the inside-out collapse model ($M=0.96$~$M_\odot$ and $R=1.6 \times 10^{17}$~cm).   This is because the ambipolar diffusion model is starting from a much earlier, more diffuse state, and a smaller fraction of the enclosed mass in the cloud would be expected to go into the protostar.  In fact, the ambipolar diffusion model is evolving towards the initial state of the CEW model, and once a sufficiently centrally condensed core has formed it will begin collapsing in a more dynamic manner.
 Figure~\ref{fig:vR_and_nR_AD} shows that this seems to happen by $\sim10^6$~yr, when there is a sharp increase in density and velocity for radii $\lesssim10^{17}$~cm.
It should also be noted that the collapse timescale is approximately one order of magnitude faster for the inside-out model than for the ambipolar diffusion model.

\section{Chemical model}
\label{sec:chemmod}

For each of the collapse models, we follow the chemical evolution of several  `test' shells spanning the radius of the core, in order to gain an estimate of the abundance profiles of various molecules at different times.  
The chemical reaction network is taken from the UMIST RATE06 database (Woodall et al. 2007), but is is limited to the elements H, C, N, O, S, He and Na.  The species list  consists of 127 gas-phase species and 40 mantle species and the reaction network contains 1816 gas-phase reactions.
The elemental abundances relative to hydrogen nuclei were taken from Rawlings \& Yates (2001) and are given in Table~\ref{EA}.
We assume that the initial conditions are atomic, apart from hydrogen, for which we assume 90\% is in the form of H$_2$.
 In Section~\ref{abundances} we also explore the effect of starting from a more evolved chemical composition in each collapse model by holding the cores static for a period of time, ranging from $10^3-10^8$~yr, before allowing them to collapse.

The chemical network of gas-phase chemical reactions includes two-body reactions, cosmic ray ionisation 
and cosmic ray induced photoreactions.  
The cosmic ray ionisation rate used is the commonly used value of  $1.30 \times 10^{-17}$~s$^{-1}$  
\citep[e.g.][]{Shem03,Garrod08}
Photoreactions resulting from the external radiation field are included in the reaction network, although we assume that the visual extinction is high ($> 10$ magnitudes), so these reactions have minimal importance.

H$_2$ formation on dust grains is included, 
the rate of which is given by the rate that hydrogen atoms stick and react to the grain surface (multiplied by 0.5 to take into account the fact that it takes two hydrogen atoms to form an H$_2$ molecule):
\begin{eqnarray}
&R_{\text{H}_2} & = \frac{1}{2} A \sqrt{3kT/m_\textrm{H}}S(\text{H})n(\text{H})  \\
&& = \frac{1}{2} \times 2.35 \times 10^{-17}  T^{1/2} n_\text{H}S(\text{H}) n(\text{H}) ~\text{cm}^{-3}~\text{s}^{-1}
\end{eqnarray}
%
%
%
 where $A$ is the grain surface area per unit volume, $T$ is the gas temperature, $n$(H) is the number density of atomic hydrogen in cm$^{-3}$, and $S$ is the sticking probability of atomic hydrogen, given by $(T/102.0 + 1)^{-2}$ \citep{Buch91}.  
 The value of $A$ is very uncertain; we have chosen a value of $1.5\times10^{-21}n_\textrm{H}$~cm$^{-1}$ (where $n_\text{H}$ is the total number density of hydrogen nuclei in cm$^{-3}$), but values can be as low as $2.6\times10^{-22}n_\textrm{H}$~cm$^{-1}$ \citep{Rawlings92}, or as high as $2.1\times10^{-21}n_\textrm{H}$~cm$^{-1}$ \citep{Duley84}.


Freeze-out of species onto grains is included, proceeding at the same rate as given in Roberts et al. (2007).  We have also included desorption of species owing to cosmic ray heating of grains at the rate given by Roberts et al. (2007), with the 
efficiency parameter $\phi$ set to $10^5$.  The value of $\phi$ is very uncertain, but using a value of $10^5$ results in the number of molecules desorbed per cosmic ray impact being consistent with the theoretical estimate of L\'eger et al. (1985) (see Roberts et al. 2007 for more detail).
We also note that other desorption mechanisms, such as desorption by H$_2$ formation on grains and desorption by the cosmic ray induced UV field may also operate in dark cores, but Roberts et al. (2007) have shown that these processes have the same qualitative effect by simply varying the value of the efficiency parameters, hence it is not necessary to include them all in this model.

Apart from H$_2$ formation, the only surface reactions that occur are hydrogenation of certain
unsaturated species, and dissociative recombination of molecular
ions. These reactions are assumed to occur instantly and the products
remain on the grain surface until they are desorbed.

The temperature is assumed to remain at 10~K throughout the collapse, and the density as a function of time for each shell is calculated by the dynamical models described in Sections~\ref{ADcoll} and \ref{CEWdescription}.

\begin{table}
\caption{Abundances of each element relative to hydrogen nuclei}
\begin{center}
\begin{tabular}{|c|c|}
H & 1.00\\
C & $1.88 \times 10^{-4}$ \\
N & $1.15\times 10^{-4}$\\
O & $6.74 \times 10^{-4}$ \\
S & $1.62 \times 10^{-7}$ \\
He & 0.1 \\
Na & $3.50 \times 10^{-7}$ \\
\end{tabular}
\end{center}
\label{EA}
\end{table}%

\section{Chemical results}
\label{chemdisc}

\subsection{Ambipolar diffusion model}
\label{sec:ADM}

\begin{figure*}
\begin{center}
\includegraphics[width=140mm]{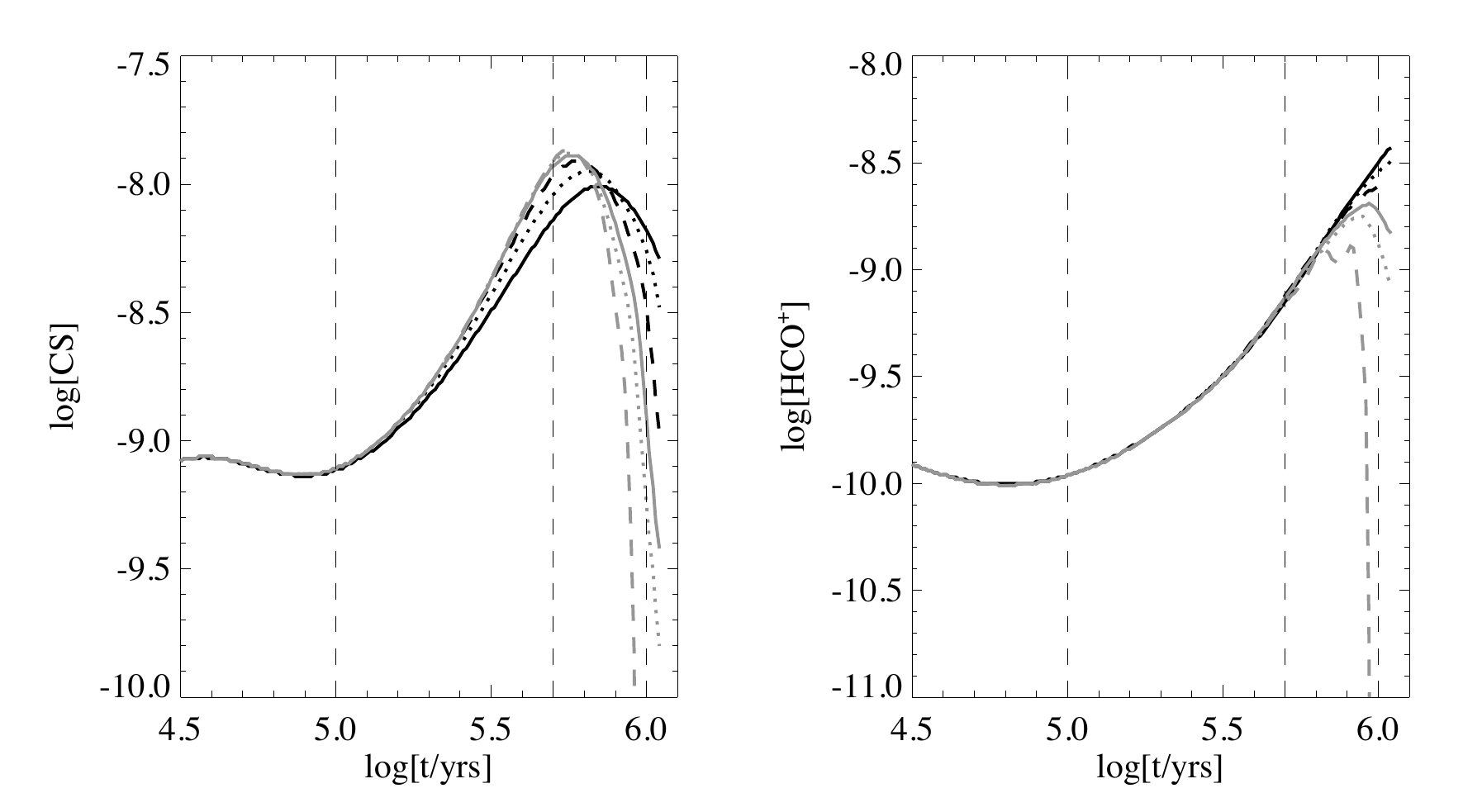}
\caption{Fractional abundances of CS and HCO$^+$ in each of the shells as a
function of time for the ambipolar diffusion model. The black solid line, black dotted line, black dashed line, grey solid line, grey dotted line and grey dashed line are for the shells beginning at $R_0 = 8.45 \times 10^{17}$, $7.73 \times 10^{17}$, $6.77 \times 10^{17}$, $5.41 \times 10^{17}$,
$3.93 \times 10^{17}$ and $1.82 \times 10^{17}$~cm respectively. The vertical dashed lines mark the times for which we
generate line profiles.}
\label{AD:ct}
\end{center}
\end{figure*}

\begin{figure*}
\begin{center}
\includegraphics[width=140mm]{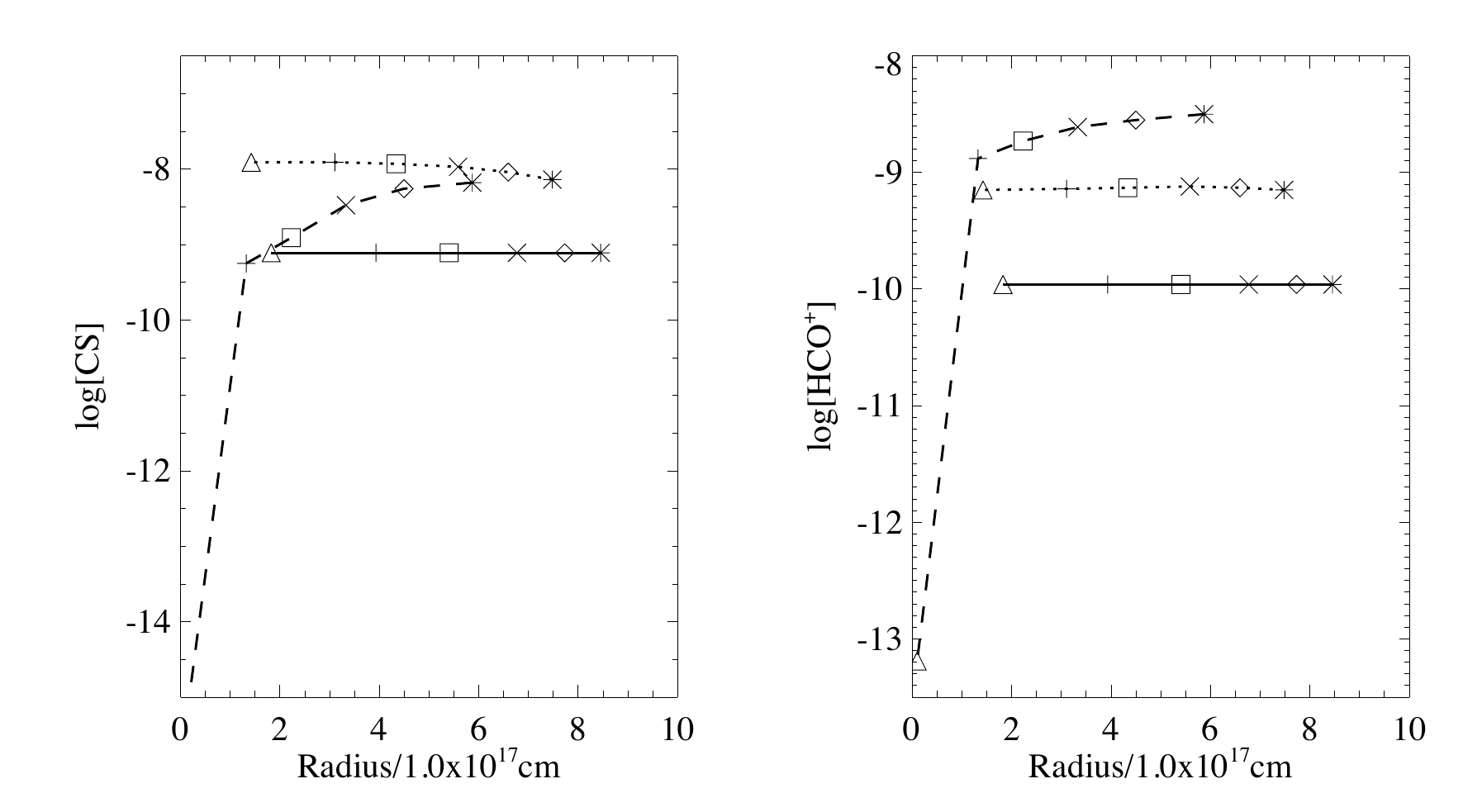}
\caption{Fractional abundances of CS and HCO$^+$ of as a function of radius for the
ambipolar diffusion model. The solid line is for $10^5$ yr, the dotted line is for $5 \times 10^5$ yr, and
the dashed line is for $10^6$ yr. The asterisks, diamonds, crosses, squares, pluses, and triangles refer to the
shells which start at 
 $R_0 = 8.45 \times 10^{17}$, $7.73 \times 10^{17}$, $6.77 \times 10^{17}$, $5.41 \times 10^{17}$,
$3.93 \times 10^{17}$ and $1.82 \times 10^{17}$~cm respectively.
}
\label{AD:cr}
\end{center}
\end{figure*}

Figure~\ref{AD:ct} shows the fractional abundances of CS and HCO$^+$ for each test shell as they evolve with time, for the ambipolar diffusion model.  These molecules were chosen because they are commonly used to search for the blue asymmetry.   Ideally, we would have liked to model H$_2$CO as well, which is another common infall tracer.  However, the main formation route of H$_2$CO is believed to be via the hydrogenation of CO on grain surfaces and subsequent desorption into the gas-phase \citep{Fuchs09}.  In our chemical model we have only included a very basic surface chemistry, which yields very low H$_2$CO fractional abundances ($\lesssim$ a few times 10$^{-10}$).  In order to obtain a more realistic H$_2$CO abundance, 
sophisticated surface chemistry would have to 
be implemented in the model, such as that employed in \citet{Cuppen09}.  In that paper they used a Monte-Carlo approach and included layering
of the ice, and found that the results were significantly different to the 
results obtained using a simple rate equation approach.  This is something that 
we can address in future work, and we therefore leave modelling the H$_2$CO line profiles until we have a suitable model.

For both CS and HCO$^+$, 
the chemical evolution  in each shell is the same up to $\sim 2 \times 10^5$~yrs.  This is because, as shown in Figure~\ref{fig:ambdiff_phys}, significant density increases only begin after this time.
 
Up until $\sim 5\times 10^5$~yrs, there is a general increase in the abundances 
as molecules form from the atomic (with the exception of molecular hydrogen) initial conditions.
After $5 \times 10^5 - 1\times 10^6$~yrs, the abundances start decreasing due to depletion.  
 However, including freeze-out in the model actually allows 
 HCO$^+$ to reach  a peak abundance a factor of 5 times higher compared to when the model is run without freeze-out, before  it  eventually depletes.
This sort of behaviour has been noted in previous studies (eg. Rawlings et al. 1992) and is complicated. It can partly be explained by considering the destruction channels of  HCO$^+$:
%
%
%
%
\begin{eqnarray}
\text{e} + \text{HCO}^+ & \to & \text{CO} + \text{H}\\
\text{H}_2\text{O} + \text{HCO}^+ & \to & \text{CO} + \text{H}_3\text{O}^+ \\
\text{C} + \text{HCO}^+ & \to & \text{CO} + \text{CH}^+.
\end{eqnarray}
%
  When freeze-out is included, the abundances of H$_2$O and C decrease as they freeze onto grains.  The electron abundance also decreases due to the depletion of easily ionised species such as C and Mg.  Therefore the destruction rate of HCO$^+$ is initially reduced by freeze-out, allowing it to reach a higher peak abundance.

Figure~\ref{AD:cr} shows the radial distribution of the fractional abundances of CS and HCO$^+$ across the core at $10^5$~yr (solid line), $5 \times 10^5$~yr (dotted line) and $10^6$~yr (dashed line), for the ambipolar diffusion model.  
These abundance profiles were generated by taking the radius of each shell at a particular time from Figure~\ref{fig:ambdiff_phys}, and taking the corresponding chemical abundance of the shells from Figure~\ref{AD:ct}.  
The plot clearly shows the contraction of the core, with the shells moving inwards as time progresses.

At $1\times 10^5$~yrs and $5\times10^5$~yrs, the abundances are relatively uniform across the core.  At $1\times 10^6$~yrs, when the collapse is well underway and there is a large density gradient across the core, both molecules suffer severe depletion for radii $\lesssim 2 \times 10^{17}$~cm.  In the outer parts of the core, the  abundance of HCO$^+$ actually increases as time progresses, due to the depletion effects described above.

\subsection{Inside-out collapse}
\label{CEWchem}

\begin{figure*}
\begin{center}
\includegraphics[width=140mm]{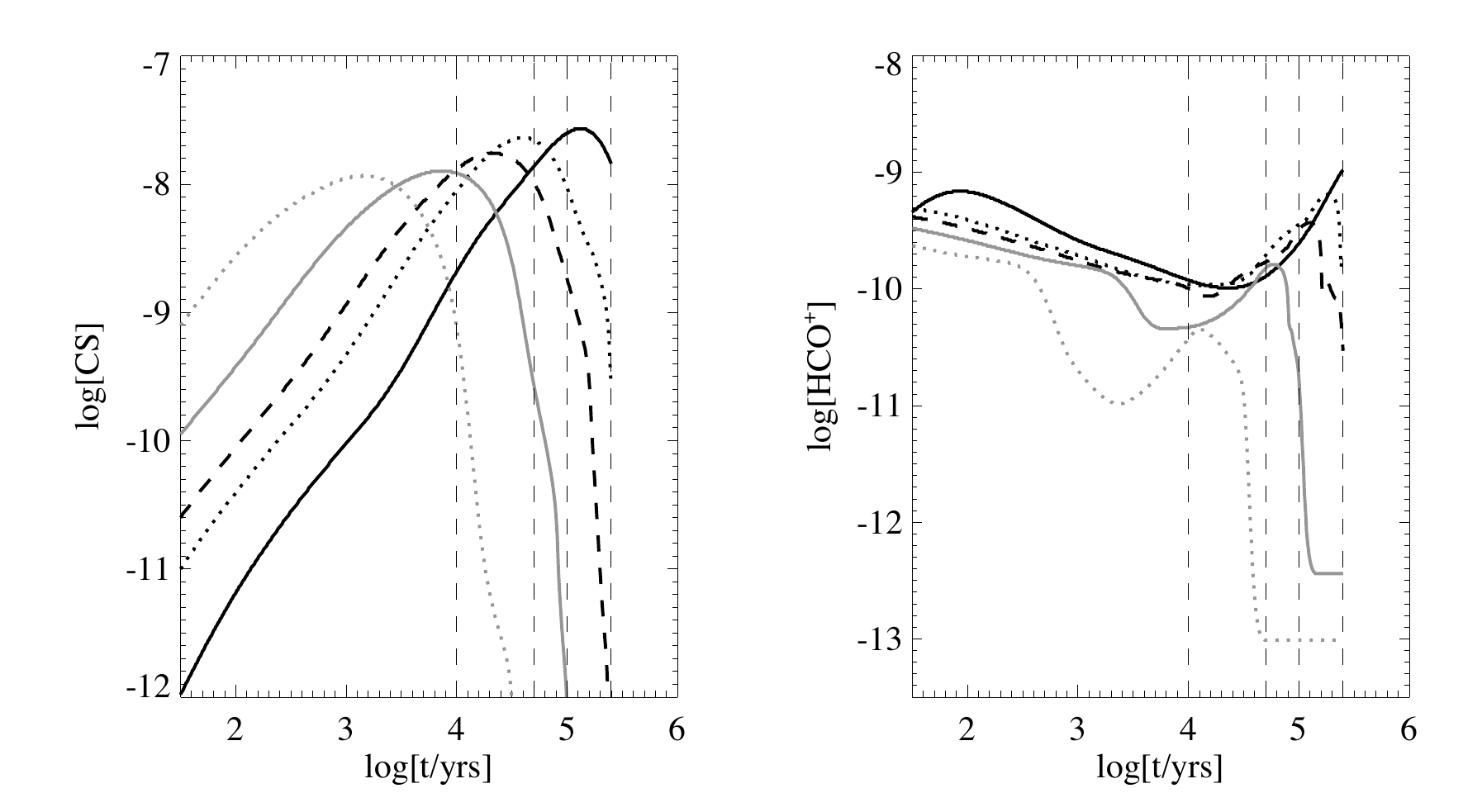}
\caption{Fractional abundances of  CS  and HCO$^+$ in each of the shells as a function of time for the inside-out collapse model. The black solid line, black dotted line, black dashed line, grey solid line and grey dotted line refer
to the shells which start at $R_0 = 1.6 \times  10^{17}$, $7.4 \times 10^{16}$, $5.0 \times 10^{16}$, $2.5 \times 10^{16}$ and $1.0 \times 10^{16}$ cm
respectively.   The vertical dashed lines mark the times for which we
generate line profiles.}
\label{CEW:ct}
\end{center}
\end{figure*}

\begin{figure*}
\begin{center}
\includegraphics[width=140mm]{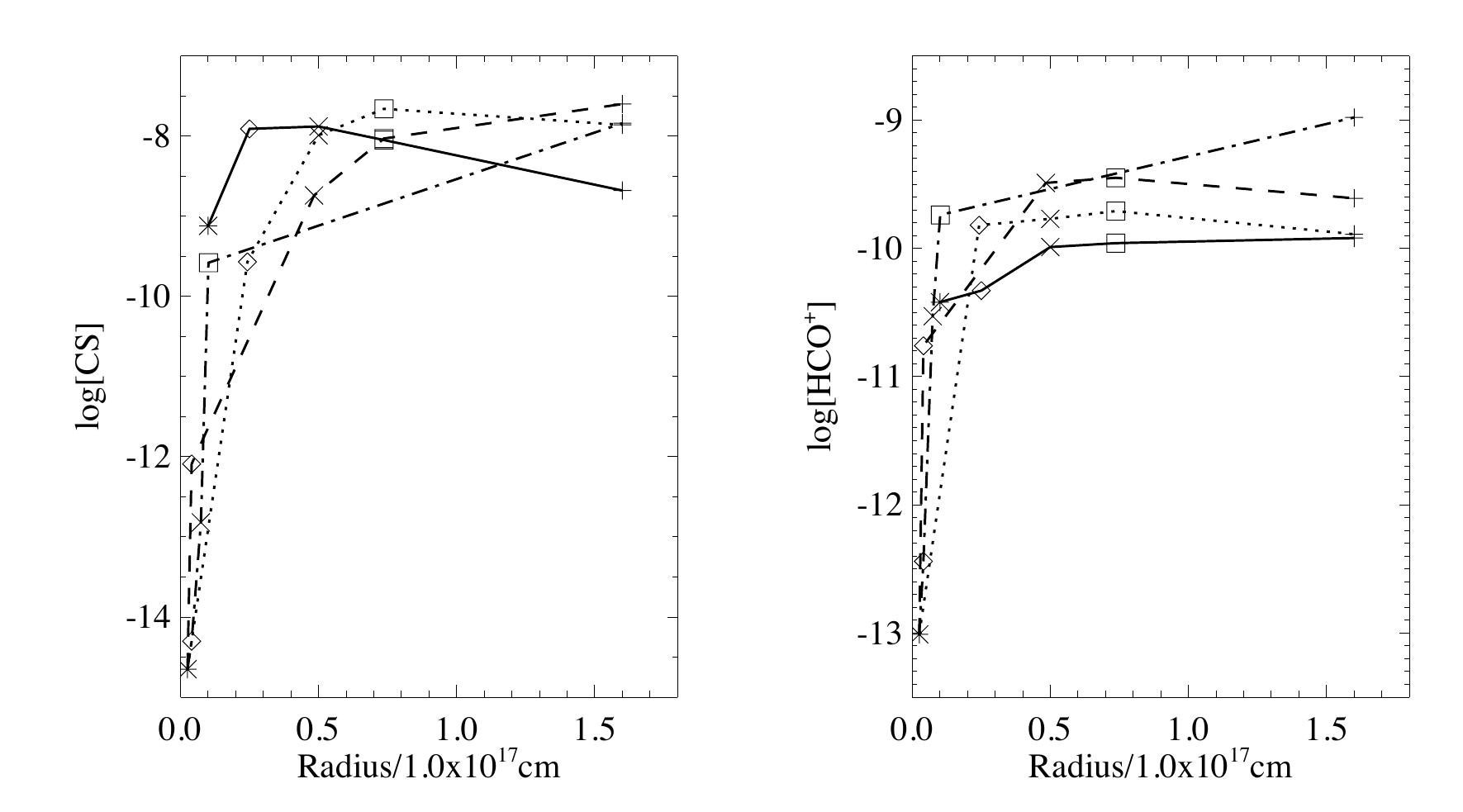}
\caption{Fractional abundances of CS and HCO$^+$ of as a function of radius for
the inside-out collapse model at different times. The solid line is for $10^4$ yr, the dotted line is for
$5\times 10^4$ yr, the dashed line is for $10^5$ yr, the dot-dash line is for $2.5\times 10^5$ yr. The pluses, squares, crosses, diamonds and asterisks represent the shells starting at
 $R_0 = 1.6 \times  10^{17}$, $7.4 \times 10^{16}$, $5.0 \times 10^{16}$, $2.5 \times 10^{16}$ and $1.0 \times 10^{16}$ cm
respectively.
}
\label{CEW:cr}
\end{center}
\end{figure*}

Figure~\ref{CEW:ct} shows the fractional abundances of CS and HCO$^+$ for each test shell as they evolve with time, for the inside-out collapse model. 
The molecules show much more complicated behaviours than for the ambipolar diffusion model for a number of reasons, such as: (i) the density of each shell actually experiences an initial {\it decrease} (see Figure~\ref{fig:cew_phys}), rather than monotonically increasing as in the ambipolar diffusion model;  
and (ii) the initial density distribution is a singular isothermal sphere (rather than a uniform sphere in the ambipolar diffusion model), so the initial conditions of the shells cover a much larger range of densities.

 For HCO$^+$  the abundance in the static outer shell (plotted in the black solid line) shows a very different evolution to the dynamically evolving inner shells.
For the inner shells, the molecules exhibit a large increase in abundance before they freeze out, due to the rapid density increase. Obviously, for outer shells, this increase in abundance occurs at later times.
The innermost shell collapses very quickly and reaches its final density of $\sim 1.1 \times 10^7$~cm$^{-3}$ after only $\sim 3.3 \times 10^4$~yr. Equilibrium between freeze-out and desorption processes rapidly follows, resulting in extremely depleted abundances.

The inclusion of freeze-out greatly increases the peak 
 abundance of HCO$^+$ before it depletes, compared to when the model is run without freeze-out
(see Section~\ref{sec:ADM}), as was found for the ambipolar diffusion model. 

Figure~\ref{CEW:cr} shows the radial abundance distribution of the same species at $10^4$, $5 \times 10^4$, $10^5$ and $2.5 \times 10^5$~yr.  
For HCO$^+$, the abundances at radii $\gtrsim 0.5 \times 10^{17}$~cm increase with time. This is due to a combination of atomic to molecular conversion and depletion enhancement effects.
Compared to the ambipolar diffusion model, the effects of depletion can be seen much earlier ($\sim 10^4$~yrs in the centre of the core),  because the dynamical timescale is much shorter. 

\subsection{Comparison to existing chemical models of collapsing cores}
 It is difficult to compare the abundances predicted by our chemical model to previously published chemical models of infalling cores, because of the varied range of parameters used (e.g. reaction data sets, elemental composition, density profiles, temperature profiles, etc.), but in this section we attempt to make a comparison with two reasonably similar studies.
  \citet{Shem03} presented the abundance profiles of a coupled dynamical and chemical model of the collapse of starless cores in magnetised molecular clouds, and found that CS and HCO$^+$ behaved in a similar manner to that which we have found in our ambipolar diffusion model; in Figures~2 and 3 of their paper they show that at the time when their model reaches a central density of $n_\textrm{H}=2\times10^6$~cm$^{-3}$ ($n(\textrm{H}_2)=10^6$~cm$^{-3}$), for their `low metal' initial abundance case (which was found to give the best fit to observations of L1544), the fractional abundances (with respect to hydrogen nuclei) of CS and HCO$^+$ were $\sim3\times10^{-8}$ and $\sim3\times10^{-9}$ respectively, which are similar (within a factor of $\sim4$) to the abundances we obtain in the outer parts of the core at a collapse time of $10^6$~yr (when the central density of the SR10 model is $\sim2\times10^7$~cm$^{-3}$).  \citet{Shem03} also find at this time that CS and HCO$^+$ are depleted in the core centre, in agreement with our model.

 \citet{Aikawa03} modelled the molecular abundances in a core collapsing according to the Larson-Penston flow \citep{Larson69,Penston69}.  Although they modelled the prestellar phase, starting from a uniform density,  the core collapses to a central density of $3\times10^7$~cm$^{-3}$ on a timescale of $2\times10^5$~yr, similar to the inside-out collapse model we have used.  As one would expect,  in our model we find that CS and HCO$^+$ deplete at an earlier time than in the \citet{Aikawa03} model; by $10^5$~yr their model shows no signs of depletion of CS or HCO$^+$, in contrast to our model at this time.  By $2\times10^5$~yr, however, the molecular profiles of CS and HCO$^+$ show similar behaviour in both models, with abundances in the outer parts of the core of $\sim10^{-8}$ and $\sim10^{-9}$ respectively, and both molecules are depleted in the core centre.

\section{Generating the line profiles}
\label{GLP}

To convert these theoretical abundance profiles into observable diagnostics we calculate the line profiles of CS and HCO$^+$. 

The transitions we consider are CS $J=5\to4$ at 244.94~GHz, CS $J=3\to2$ at 146.97~GHz, CS $J=2\to1$ at 97.98~GHz, 
and HCO$^+$ $J=3\to2$ at 267.56~GHz, and HCO$^+$ $J=1\to0$ at 89.19~GHz.  These transitions were chosen because they are usually strong and are good tracers of the kinematics in protostellar infall sources (Zhou et al. 1993; Evans et al. 2005).

\begin{figure}
\begin{center}
\includegraphics[width=90mm]{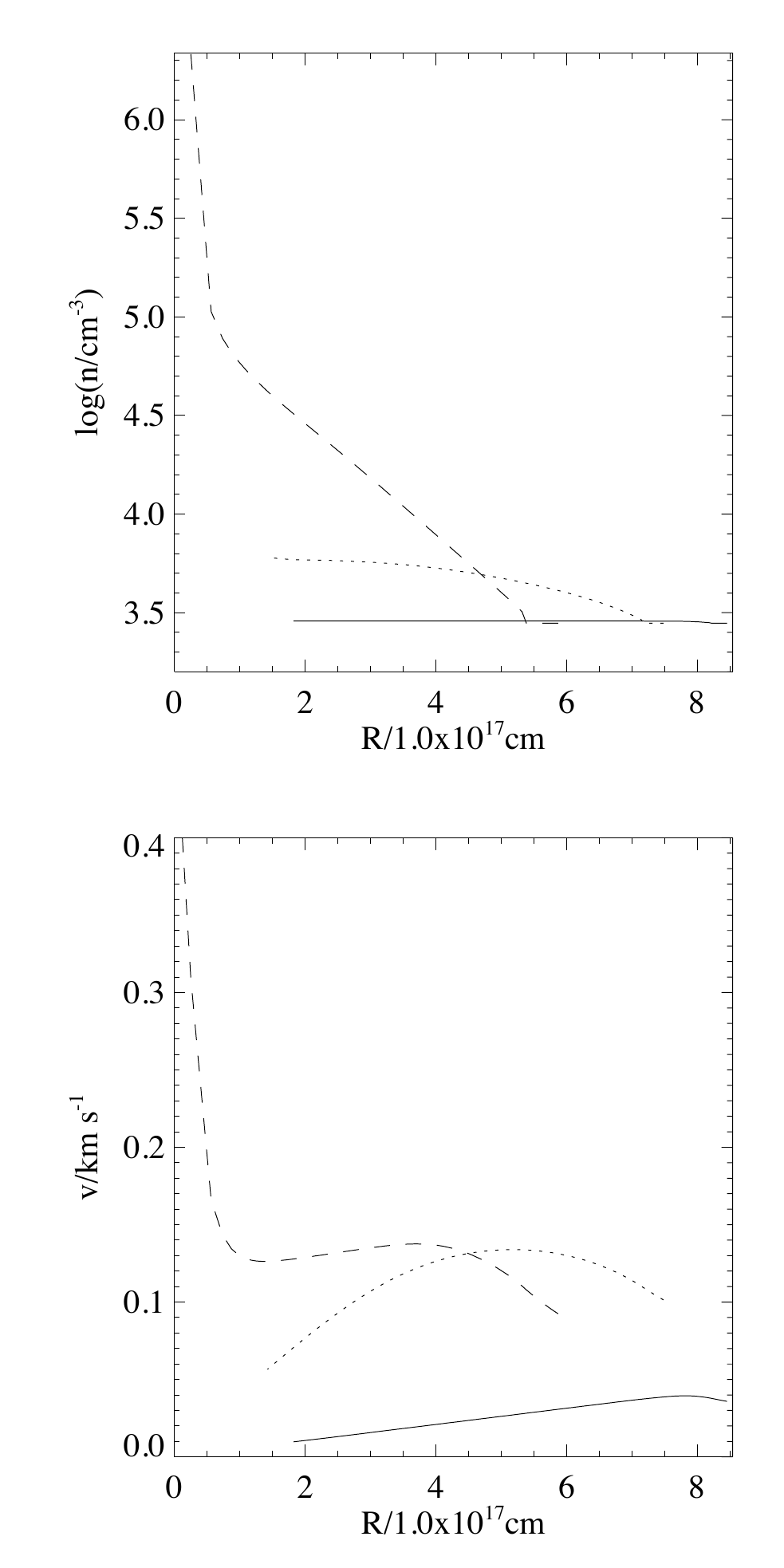}
\caption{Density and velocity profiles for the ambipolar diffusion model at $1 \times 10^5$~yr (solid line), $5 \times 10^5$~yr (dotted line) and $1 \times 10^6$~yr (dashed line).}
\label{fig:vR_and_nR_AD}
\end{center}
\end{figure}

\begin{figure}
\begin{center}
\includegraphics[width=90mm]{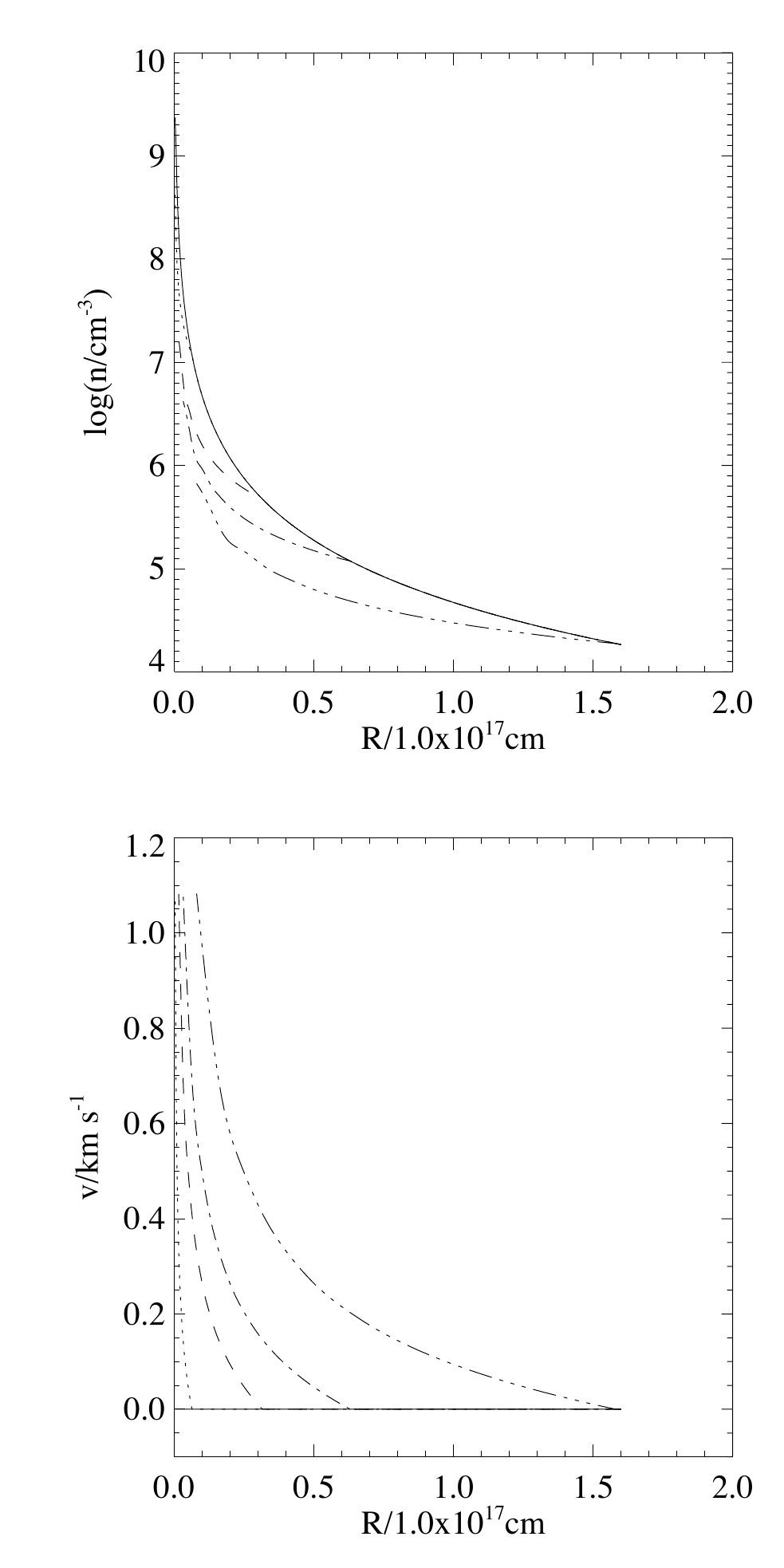}
\caption{Density and velocity profiles for the inside-out collapse model at $t=0$~yr (solid line), $10^4$~yr (dotted line), $5 \times 10^4$~yr (dashed line) $1 \times 10^5$~yr (dot-dashed line) and $2.5\times 10^5$~yr (triple dot-dashed line).}
\label{fig:vR_and_nR}
\end{center}
\end{figure}

The spherical radiative transfer code SMMOL (`Spherical Multi Mol') was used to generate the line profiles (Rawlings \& Yates 2001).  This is an approximate lambda-iteration code, which can be used for radiatively coupled clouds (as is likely to be the case for the early stages of core collapse), where the standard large velocity gradient (LVG) approximation is not appropriate.  SMMOL was benchmarked for accuracy by  \citet{vZad02}.
At each radial point in the core, SMMOL calculates the level populations,
the line source functions and the emergent line profile.
SMMOL requires input tables for each molecule giving the density, fractional abundance, gas temperature, dust temperature, infall velocity and turbulent velocity at different radii for a particular time.  To generate these tables, we took 100 evenly spaced values of the radius between the innermost and outermost shells, and to obtain the abundances at each radii we interpolated between the data shown in Figures~\ref{AD:cr} and \ref{CEW:cr}. 
We tested that this was sufficient resolution by running the model with various grid sizes, and found that the spectra were invariant for $\gtrsim 70$ radial grid points.  Following Rawlings \& Yates (2001), we used a convergence criterion of $\Delta n_i/n_i \leq 10^{-4}$ (where $n_i$ is the population of level $i$).
As a boundary condition SMMOL assumes that the material within the inner radius is static and at a constant abundance. For the densities that we are considering, we make the reasonable assumption that the gas and dust are thermally coupled and have equal temperatures\footnote{
 This assumption may be inaccurate for the early stages of the ambipolar diffusion model, where the density is $<10^4$~cm$^{-3}$ throughout the core.  However, at dust temperatures of 10--20~K, the dust emission only dominates the background continuum for frequencies $\gtrsim100$~GHz so  the lower frequency lines should be unaffected by this assumption. 
}.  
We have used a constant temperature of 10~K throughout the core, which has been found to be a valid approximation for pre-stellar cores such as L1498 and L1517B (Tafalla et al. 2004), although we also investigate other temperature profiles. 
For the background radiation field, we have used the standard interstellar radiation field of 
\citet{Mathis83}, 
supplemented by a black body at 2.7~K for the cosmic background radiation. 
We have used a constant turbulent velocity of 0.12~km~s$^{-1}$, which was 
 the 1/e velocity width that \citet{Zhou90} and \citet{Zhou93} found to give the correct velocity width for observations towards B335.
%
The velocity and density profiles for the ambipolar diffusion and the inside-out collapse  models are shown in Figures~\ref{fig:vR_and_nR_AD} and~\ref{fig:vR_and_nR}.

The cores were assumed to be at a distance of 140~pc, the distance of the Taurus Molecular Cloud.  For the telescope beam we have used the characteristics of the IRAM 30m,  centred on the source.  These are listed in Table~\ref{telescope}.  At this distance the cores have angular sizes of $\sim 800''$ and $\sim 150''$ for the ambipolar diffusion model and inside-out model respectively, so are $\sim 30$ times and $\sim 5$ times larger than the largest beam size (which is $27.5''$ at 89~GHz).  

\begin{table}
\caption{Characteristics of the IRAM 30m telescope at the frequencies of the lines we are generating, taken from the IRAM website (http://www.iram.fr/IRAMES/index.htm).  HPBW is the half-power beam width and $\eta_\text{a}$ is the aperture efficiency.
}
\begin{center}
\begin{tabular}{ccc} 
Frequency (GHz) & HPBW (arcseconds) & $\eta_\text{a}$ \\ \hline \hline
89 & 27.5 & 0.61 \\
98 & 25 & 0.60 \\
141 & 17.5 & 0.55 \\
147 & 16.5 & 0.55 \\
226 & 11 & 0.42 \\
245 & 10 & 0.39 \\
268 & 9 & 0.36 \\ 
\end{tabular}
\end{center}
\label{telescope}
\end{table}%

\section{Results and discussion}
\label{RAD}

\subsection{Ambipolar diffusion line profiles}
\label{ADprofiles}

\begin{figure*}
\begin{center}
\includegraphics[width=140mm]{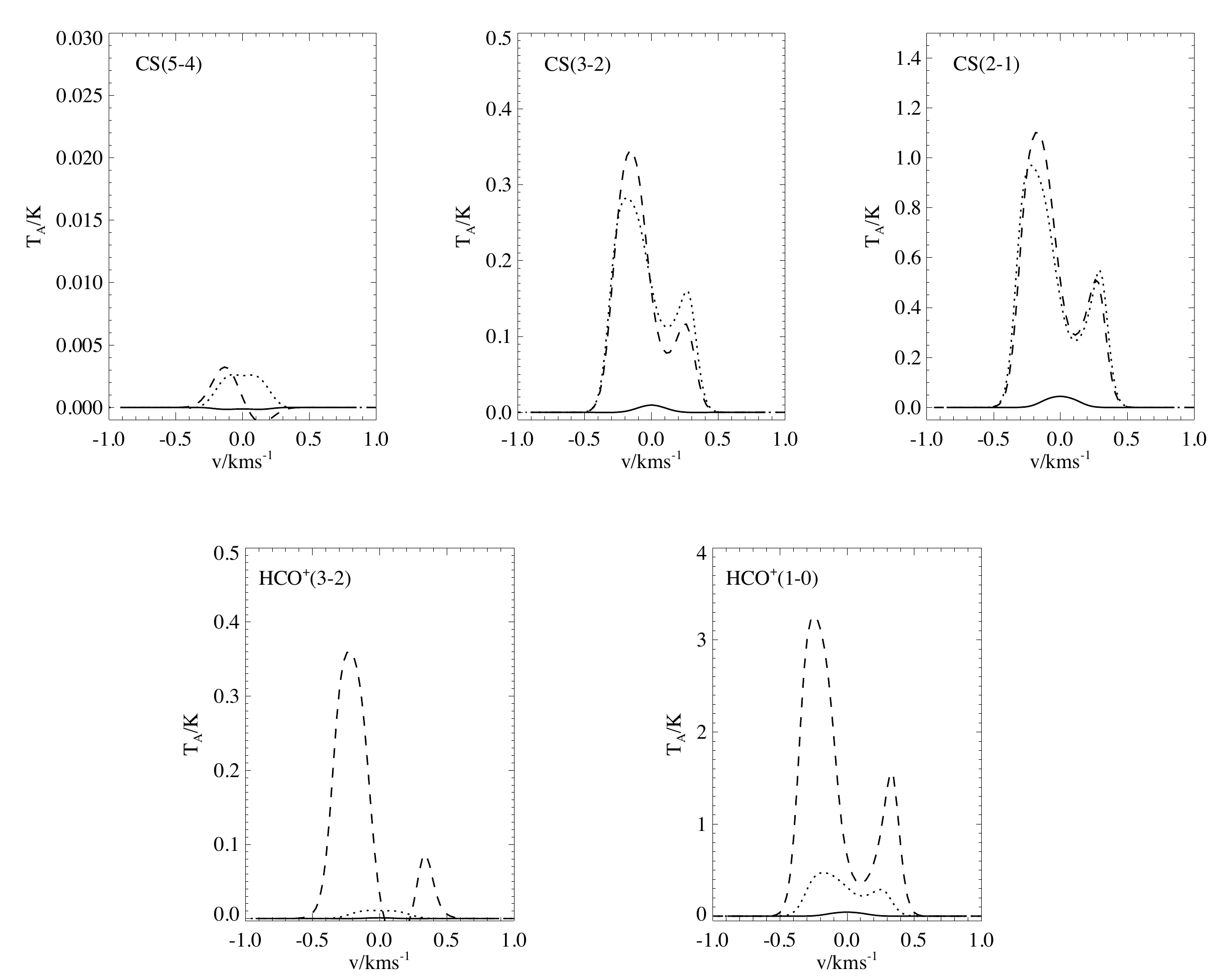}
\caption{Line profiles of CS for transitions $J = 5 \to 4$, $3 \to 2$, and $2 \to 1$,  HCO$^+$ $J = 3 \to 2$ and $1 \to 0$  for the ambipolar
diffusion model at $1 \times 10^5$ yr (solid line), $5 \times 10^5$ yr (dotted line) and $1 \times 10^6$ yr (dashed line).}
\label{all_AD}
\end{center}
\end{figure*}



The line profiles for CS $J=5\to4$, $3\to2$, $2\to1$ and  HCO$^+$ $J=3\to2$ and $1\to0$,
are shown in Figure~\ref{all_AD},
for  collapse times of $1 \times 10^5$~yr, $5 \times 10^5$~yr and $1 \times 10^6$~yr.

In this model the thermal velocity is $v_\text{th} = \sqrt{kT/m(i)} \sim 0.043$~km~s$^{-1}$ for CS at 10~K.  At $1 \times 10^5$~yr the maximum infall velocity is 0.038~km~s$^{-1}$, meaning that the line-width is dominated by turbulence and the core is essentially radiatively coupled.  At $5 \times 10^5$~yr and $1 \times 10^6$~yr, however, the maximum infall velocity is 0.12~km~s$^{-1}$ and 0.31~km~s$^{-1}$ respectively. The systemic infall therefore begins to dominate the line-broadening, and at $10^6$~yr the core centre and the core edge are radiatively decoupled  because of the relatively large velocity gradient. 


 Our model predicts that the CS $J=5\to4$ lines are undetectably weak, and that the $3\to2$ and $2\to1$ lines only become detectable for times $\ge 5\times 10^5$~yr, with maximum intensities of $\sim 0.4$ and $1.0$~K.  This increase in intensity at these times is due to the increase in density of the core.
The line profiles for  CS $J=3\to2$ and $2\to1$ are very strongly self-absorbed and asymmetric. By $5 \times 10^5$~yr these transitions have a noticeably asymmetric profile with a stronger blue-shifted  peak.
It can be seen that the absorption `dip' is off-centre, and it is most redshifted at $5 \times 10^5$~yr, where the minimum lies at $\sim 0.1$~km~s$^{-1}$, which is approximately the velocity of the material in the outer regions of the core at this time (see Figure~\ref{fig:vR_and_nR_AD}).
Unlike the inside-out collapse, where the envelope material is static,
this redshift in the absorption is the main cause of the blue-skewed asymmetry in the line-profiles. 
Such off-centre absorption dips have been observed towards the class 0 protostar, IRAM 04191, where they are also believed to be caused by extended inward motions (Belloche et al. 2002).

In the case of HCO$^+$, the $J=3\to2$ line only becomes detectable at $10^6$~yrs.  The $1\to0$ line is detectable by $5\times 10^5$~yrs, and its peak intensity strongly increases by a factor of $\sim 7$ by $10^6$~yrs.  This is due to the increase  in gas density and in abundance of HCO$^+$ for radii $\gtrsim 2 \times 10^{17}$~cm (see Figure~\ref{AD:cr})
By $10^6$~yr, both the $3\to2$ and $1\to0$ lines are very strongly self absorbed with a clear blue asymmetry.


 \citet{Pav03}  also modelled CS and HCO$^+$ line profiles in a spherically symmetric, coupled dynamical and chemical model of a prestellar core whose evolution is determined by ambipolar diffusion.  Their model core was tailored to fit the stareless core L1544, and had a much larger mass (20~M$_\odot$) and radius ($10^5$~AU $\sim1.5\times10^{18}$~cm) than the model of SR10.  
  Due to the large differences core sizes and therefore molecular column densities, we cannot make a quantitative comparison to their results, but 
 nevertheless, their resulting line profiles show notable qualitative similarities to the ones we have derived.  In Figure~2 of \citet{Pav03}, they present their HCO$^+$($3\to2$) and CS($2\to1$) line profiles for 6.5~Myr, when the  central density is $n_\text{H}\sim2\times10^6$~cm$^{-3}$.
Comparing these to our Figure~\ref{all_AD}, for a collapse time of 1~Myr (by which time the central density has reached $\sim10^7$~cm$^{-3}$), we see that both models predict a very strong blue asymmetry in both lines, and  a red-shifted abosrption dip.  The main difference with the results of \citet{Pav03} is that the 
absorption dips are somewhat wider, so the  red-peaks therefore lie at a higher velocity ($\sim0.5$~km~s$^{-1}$) than we have found ($0.2-0.3$~km~s$^{-1}$).
%
Furthermore, the CS($2\to1$) line in the model of \citet{Pav03} has a much higher blue-to-red peak intensity ratio; they predict a ratio of $\sim4$, whereas we find it only to be 2.2.  These differences appear to be because the infall velocity in the core envelope of the \citet{Pav03} model reaches higher velocties ($\sim0.2$~km~s$^{-1}$) compared to those in SR10's model ($\sim0.14$~km~s$^{-1}$).  SR10's model reaches a higher velocity in the core centre ($\sim0.4$~km~s$^{-1}$), but the molecules are strongly depleted in this region so it does not have much effect on the line profiles.

\citet{Pav03} noted that their model predicted too high infall velocities to fit the observations of L1544 by \citet{Lee01} and \citet{Lee04}.  In this respect the model of SR10 appears to perform slightly better,  despite the fact that the mass-to-flux ratios are similar in both models ($\sim2$ and $\sim 3$ times the critical value in the models of \citet{Pav03} and SR10 respectively). 
The differences in infall velocities could be due to differences in boundary conditions, and the fact that \citet{Pav03} do not take into account the drag force exerted by charged dust grains.


\subsubsection{Varying the  kinetic temperature profiles for the ambipolar diffusion collapse}
\label{temp_AD}

\begin{figure}
\begin{center}
\includegraphics[width=90mm]{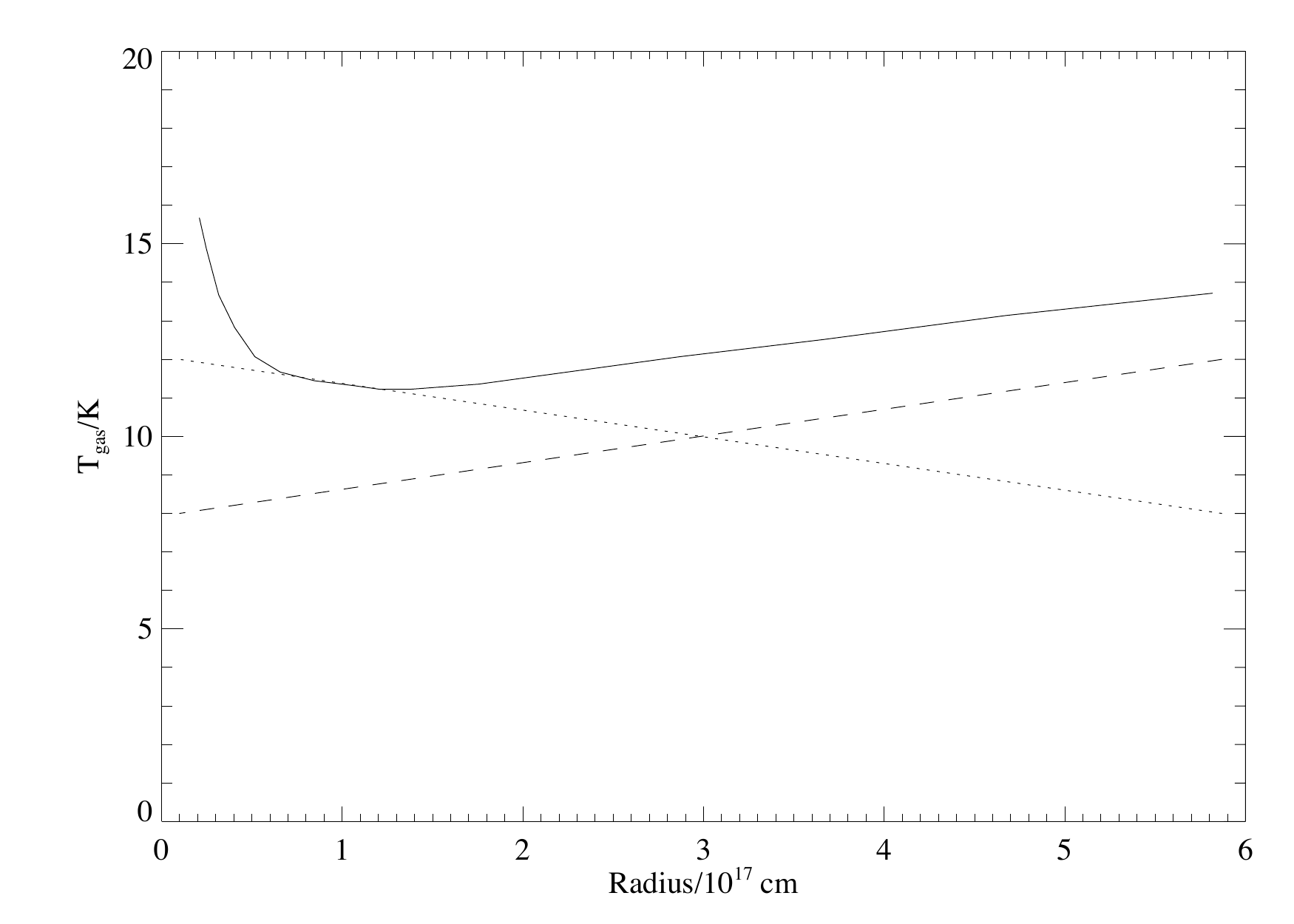}
\caption{The temperature profiles tested.  The solid line shows the temperature profile of the Class 0 source B335 (Zhou et al. 1990), normalised to a radius of $5.9\times10^{17}$~cm.  
 The dotted and dashed lines show the two linear temperature profiles.
}
\label{tempprofiles}
\end{center}
\end{figure}

\begin{figure*}
\begin{center}
\includegraphics[width=140mm]{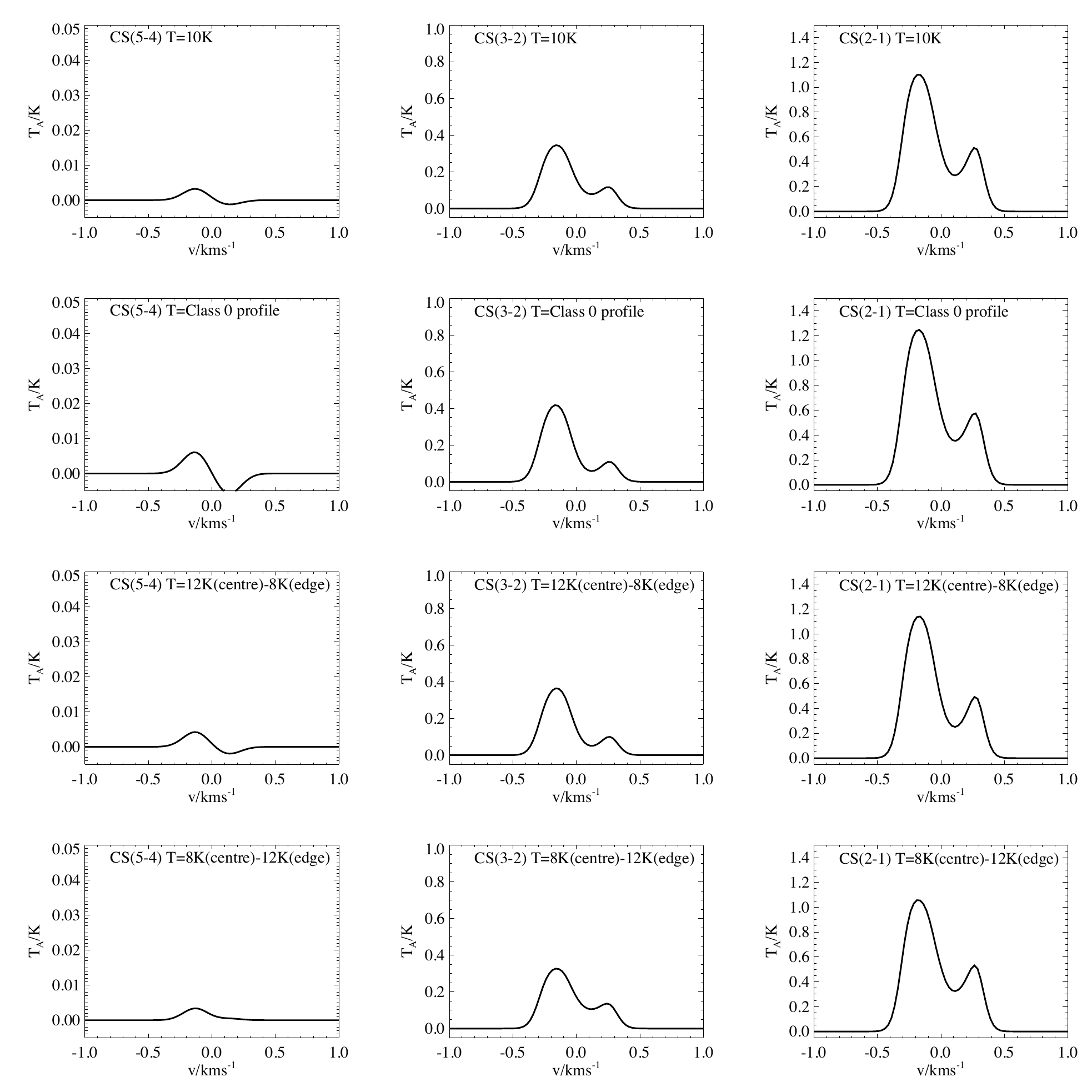}
\caption{Plot of CS spectra at $1 \times 10^6$~yr for different temperature profiles, for ambipolar diffusion collapse.   
Top row: constant temperature of 10K; second row: Class 0 temperature profile;  third row: linear temperature profile which is 12~K in the centre and 8~K at the edge; bottom row: linear temperature profile which is 8~K in the centre and 12~K at the edge. }
\label{fig:ambdiff_temp}
\end{center}
\end{figure*}

\begin{table}
\caption{The ratio of blue-to-red peak intensities for different temperature profiles for the ambipolar diffusion model at $1 \times 10^6$~yr.  The $5\to4$ line has not been included because it does not have an obvious double peak structure.}
\begin{center}
\begin{tabular}{l|ccc}
Temperature profile & CS($3\to2$) & CS($2\to1$) \\ \hline \hline
10~K &2.97 & 2.15\\
Class 0 &3.85 & 2.17\\
12K centre - 8K edge  & 3.63  & 2.31 \\
8K centre - 12K edge  & 2.40  & 1.99 \\
\end{tabular}
\end{center}
\label{bluepeaks2}
\end{table}

So far, we have assumed a constant temperature of 10~K across the core, but now we look at how the line profile shapes are affected by other temperature profiles to test the sensitivity and possible degeneracies of the results to the free parameters.

The first temperature profile we consider is that deduced from dust continuum observations of the Class 0 source B335 \citep{Zhou90}. The shape of this profile comes from heating by the internal protostar in the centre, and heating by the interstellar radiation field at the edge of the core.  
 Although the ambipolar diffusion model mostly describes the pre-stellar phase, by $10^6$~yr the conditions are similar to the protostellar regime and therefore the class 0 temperature profile may be more appropriate than the constant temperature profile at this stage.
The radius has been normalised to  $5.9 \times 10^{17}$~cm, to fit the radius of our model core at a collapse time of $10^6$~yr. 
Also, to understand more about the behaviour of the line profiles, we have tested two ad hoc linear temperature profiles, one of which is warm in the centre (12~K) and cooler on the outside (8~K), and the other of which is cold (8~K) in the centre and warmer (12~K) on the outside.  The temperature profiles are shown in Figure~\ref{tempprofiles}. 

The results are shown in Figure~\ref{fig:ambdiff_temp} for the CS $J=5\to4$, $3\to2$ and $2\to1$ lines at $1\times 10^6$~yr.  
 The overall characteristics of the  $3\to2$ and $2\to1$ line profiles do not change (i.e. a strong blue asymmetry with an off-centre absorption dip), although the ratios of the blue-to-red peak intensities (given in Table~\ref{bluepeaks2}) are relatively sensitive to the temperature profiles.
 For the Class 0 temperature profile and the linear temperature profile which is 12~K in the centre, the $5\to4$ line shows some absorption against the background continuum.  The intensities are extremely low, however, so this feature would not be observable.

\subsection{Inside-out line profiles}

\begin{figure*}
\begin{center}
\includegraphics[width=140mm]{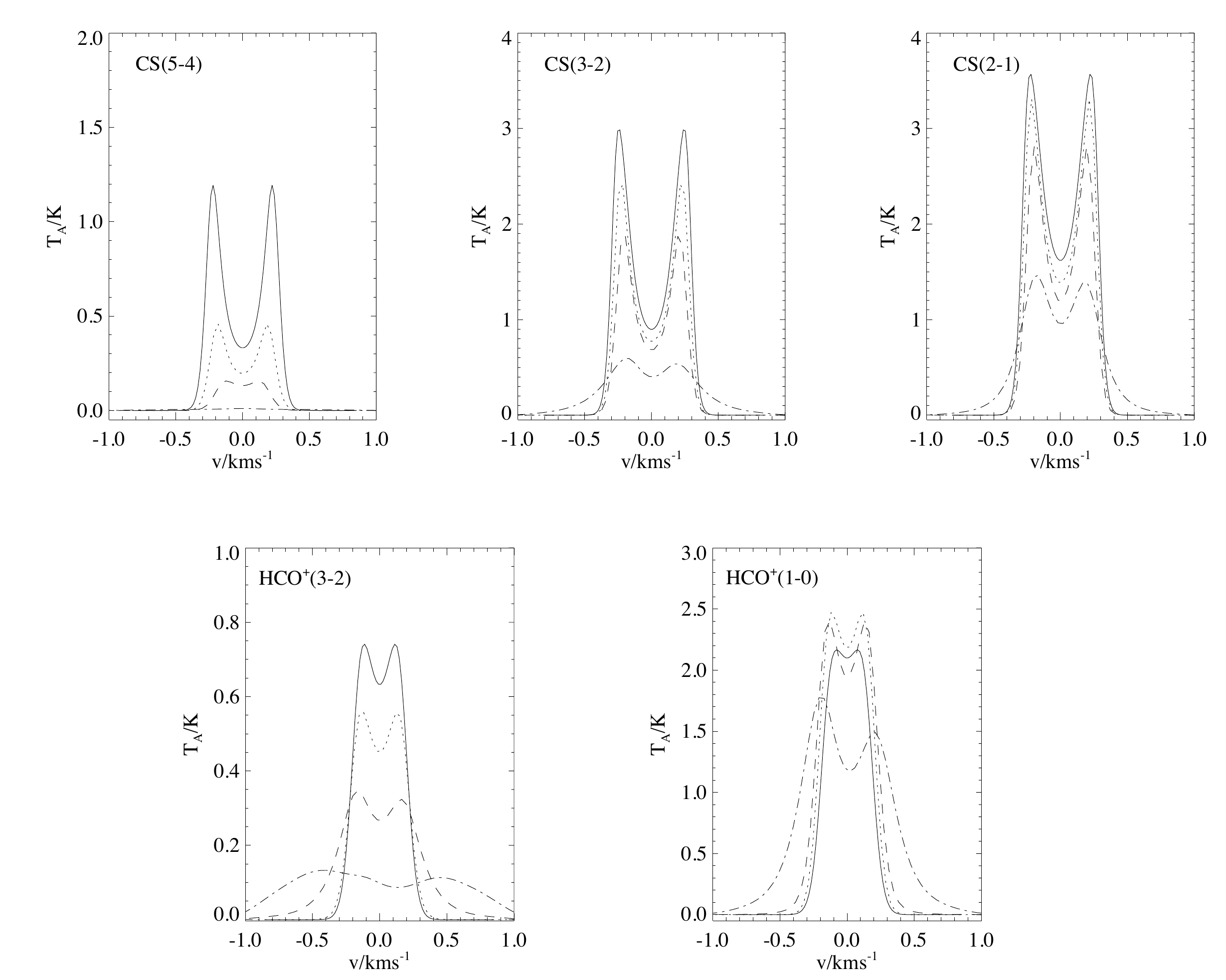}
\caption{Line profiles of CS for transitions $J = 5 \to 4$, $3 \to 2$, and $2 \to 1$, HCO$^+$ $J = 3 \to 2$ and $1 \to 0$ for the inside-out collapse model at 
 $1 \times 10^4$ yr (solid line), $5 \times 10^4$ yr (dotted line), $1 \times 10^5$ yr (dashed line) and $2.5 \times 10^5$ yr (dot-dashed line).
}
\label{all_CEW}
\end{center}
\end{figure*}



Line profiles are plotted for the same transitions as in Section~\ref{ADprofiles}, for  the inside-out collapse model at $10^4$, $5 \times 10^4$, $10^5$ and $2.5 \times 10^5$~yr.  These are shown in Figure~\ref{all_CEW}.

For $t\ge 10^4$~yr, the inner material in the core reaches velocities of $\sim 1.0$~km~s$^{-1}$, but the outer material remains static, meaning that the central regions of the core become largely radiatively decoupled from the outer static envelope.

The line profiles for CS, show double peaked profiles that are approximately symmetric until the final time, $2.5 \times 10^5$~yr. This asymmetry develops because once the CEW has reached the boundary of the core there is no longer a static outer envelope. The outer material responsible for the absorption is now redshifted as it falls towards the core, and the profile kinematics are then similar to what was seen in the ambipolar diffusion case: the minimum in the absorption dip is slightly redshifted at this time (only), at $\sim 0.02$~km~s$^{-1}$.
Thus, it would appear that if there is no imposed kinetic temperature gradient then, regardless of the nature of the collapse, the presence of a blue asymmetry is indicative of an infalling outer envelope.

The peak intensities of all three CS lines decrease with time. Referring to Figure~\ref{CEW:cr}, this appears to be a consequence of the decrease in CS abundance in the central regions of the core due to freeze-out,  as well as the decrease in density (at a fixed radius) which is shown in Figure~\ref{fig:vR_and_nR}, which affects the collisional excitation as well as decreasing the CS column density.

 For early collapse times ($\leq 1\times10^5$~yr), the CS line intensities for the inside-out collapse model are much higher than for the ambipolar diffusion model, because at this stage the inside-out collapse model has much higher densities and there is still a relatively high CS abundance in the core centre.
 By $2\times10^5$~yr in the inside-out collapse model and $1\times10^6$~yr in the ambipolar diffusion model, the central core densities and CS depletion are similar in the two models, which is reflected in the similar CS line intensities at these times. 


The HCO$^+$ line profiles are strong, symmetric and double peaked.
A blue asymmetry develops at later times; by $10^5$~yr, the $3\to2$ line has a marked asymmetry and enhanced line width. At $2.5\times10^5$~yr both lines are blue-asymmetric.
Whilst the $1\to0$ line line strength is reasonably constant (except at late times), that of the $3\to2$ line declines steadily with time.
 Again, as a result of the higher densities, 
the HCO$^+$ line intensities are much greater in the inside-out collapse model compared to the ambipolar diffusion model (apart from when the ambipolar diffusion model reaches $10^6$~yr,  when it also has a high central density). 


\subsubsection{Varying the  kinetic temperature profiles for the inside-out collapse}
\label{CEW-temp}

In the context of the inside-out collapse model (with a static outer envelope) it is usually stated that, in order to have asymmetric line profiles with a stronger blue-shifted peak, a negative excitation temperature gradient is needed (Zhou et al. 1993).

\begin{figure}
\begin{center}
\includegraphics[width=85mm]{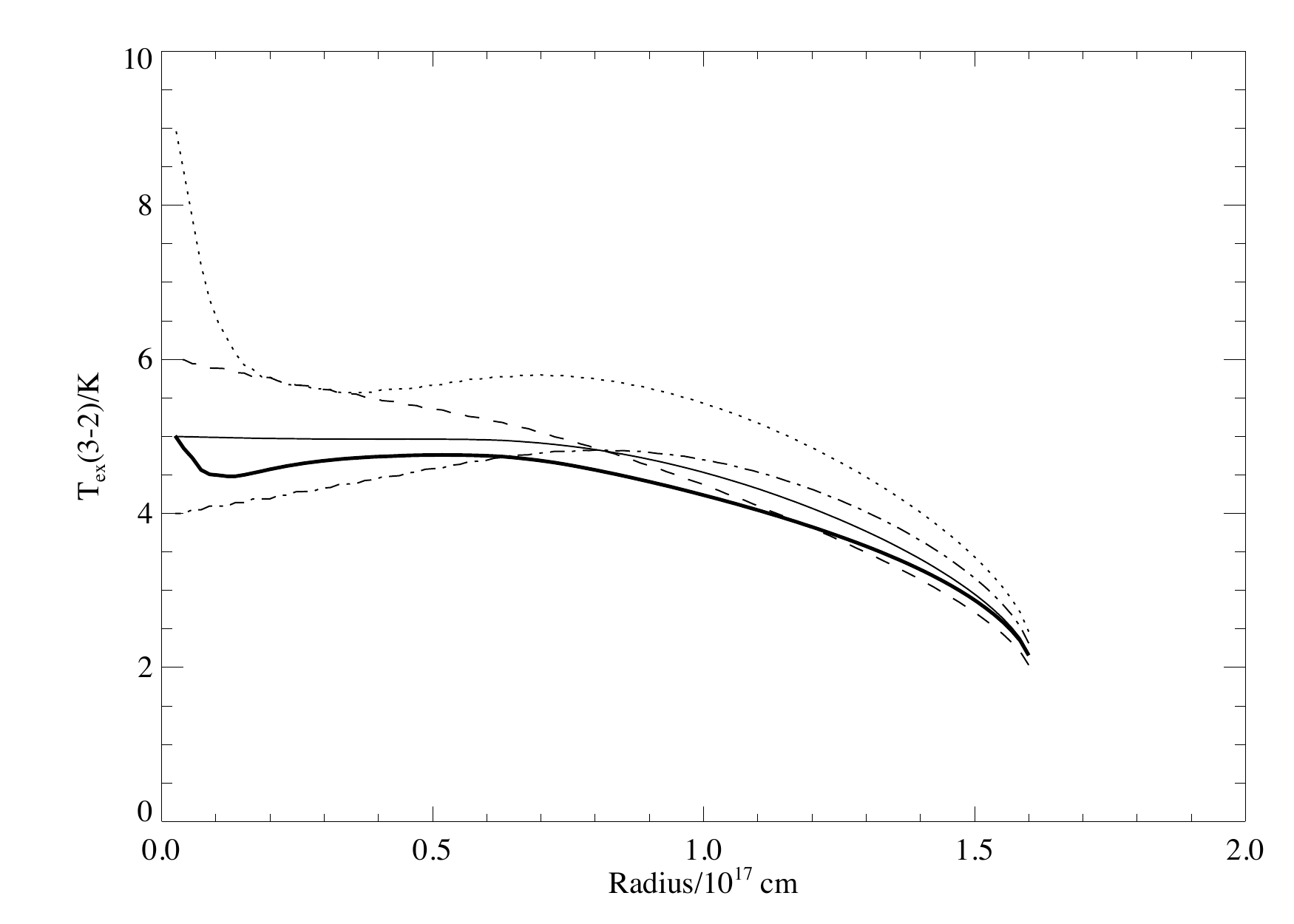}
\caption{The excitation temperature, $T_\textrm{ex}$, of the CS($3\to2$) line as a function of radius, for the inside-out collapse model at a collapse time of 10$^5$~yr.  The thick solid line shows $T_\textrm{ex}$ for a constant kinetic temperature profile of 10~K, using the CS abundance profile of Figure~\ref{CEW:cr}.  The thin lines show $T_\textrm{ex}$ using a constant CS abundance of $2.5\times10^{-8}$, with a constant kinetic temperature gradient of 10~K (thin solid line), a Class 0-type temperature profile (dotted line), a linear temperature profile with 12~K in the core centre and 8~K at the core edge (dashed line), and a linear temperature profile with 8~K in the core centre and 12~K at the core edge (dot-dashed line).}
\label{Tex}
\end{center}
\end{figure}

 In Figure~\ref{Tex}, the thick solid line plots the excitation temperature, $T_\textrm{ex}$, of the CS($3\to2$) line for the spectrum plotted in Figure~\ref{all_CEW} for a collapse time of $10^5$~yr.
At this time, there is  a negative excitation gradient for radii $\geq 5.2\times10^{16}$~cm, increasing from 2.2~K at the outer radius to 4.8~K at $5.2\times10^{16}$~cm.  For radii of $1.2\times10^{16}<R<5.2\times10^{16}$~cm,  however, there is a positive excitation gradient.  The infall radius at this time is $\sim6\times10^{16}$~cm, so the infalling material is within the region with the positive excitation temperature gradient.  Therefore, the central infalling blue-shifted material does not have an excitation temperature higher than the red-shifted infalling material at the edge of the collapse expansion wave, and this can explain why we do not see the blue asymmetry at this time.  We note that for radii $<1.2\times10^{16}$~cm, the excitation temperature sharply increases towards the core centre, but for such small radii the CS is heavily depleted.

If a negative kinetic temperature gradient is imposed, this can change the excitation temperature gradient in the core, and could give rise to the blue asymmetry. 
%
However, we also found that the CS line profiles remained symmetric for all times 
$\leq 1 \times 10^5$~yr even when adopting the various temperature profiles described in Section~\ref{temp_AD}.
Rawlings \& Yates (2001) also found that CS exhibited symmetric line profiles for inside-out collapse, and attributed this to the depletion of CS in the central regions.

\begin{figure*}
\begin{center}
\includegraphics[width=140mm]{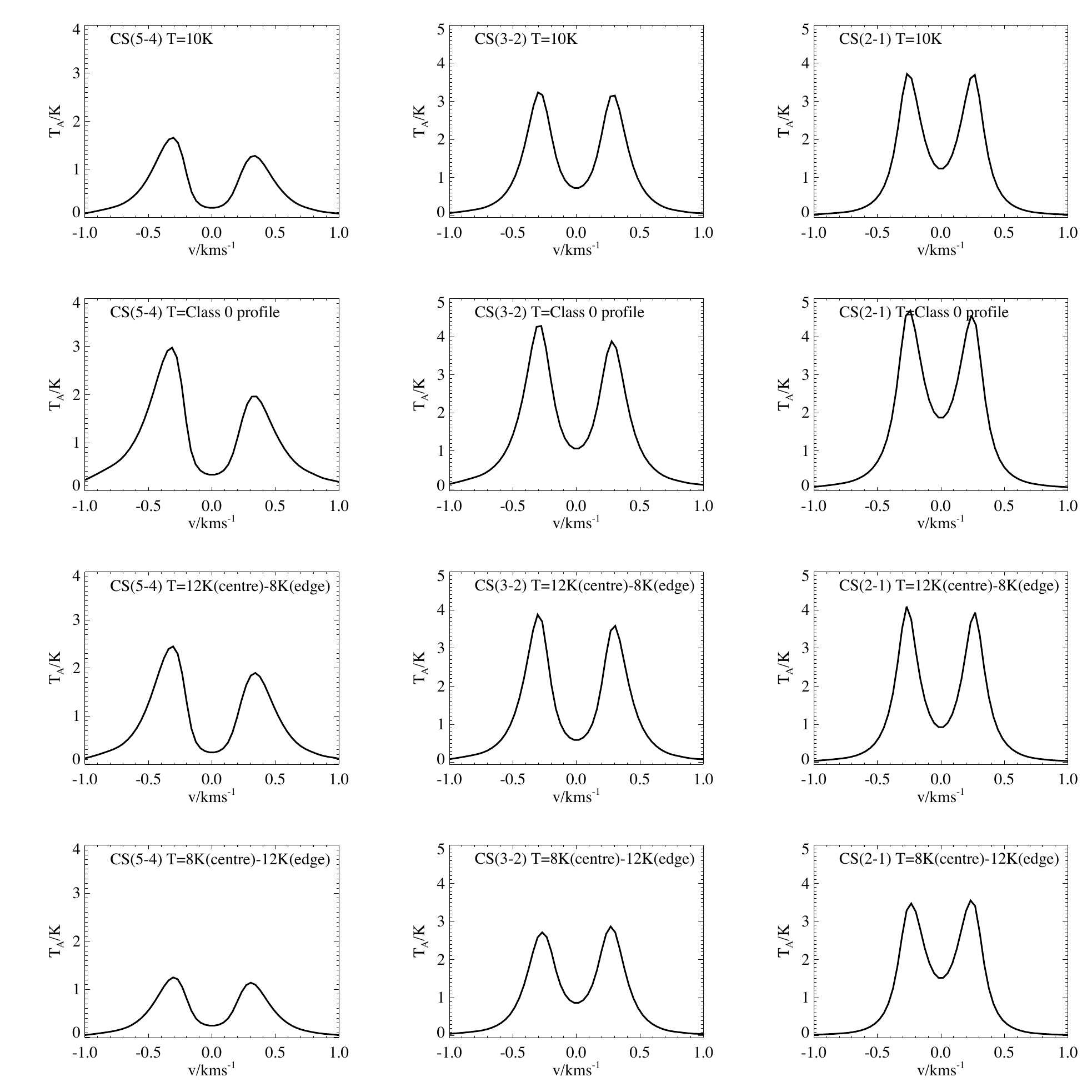}
\caption{Plot of CS spectra for $1.0 \times 10^5$ yr for different temperature profiles, for inside-out collapse assuming a constant CS abundance across the core. Top row: constant temperature (10K); second row: Class 0 temperature profile;  third row: linear temperature profile which is 12~K in the centre and 8~K at the edge; bottom row: linear temperature profile which is 8~K in the centre and 12~K at the edge.}
\label{fig:cew_temp}
\end{center}
\end{figure*}

To test this theory, we have calculated the line profiles for CS at $10^5$~yr using the temperature profiles  described in Section~\ref{temp_AD}, but this time assuming a constant fractional abundance of CS across the core.  We have used a fractional abundance for CS of $2.5\times 10^{-8}$, which is the CS abundance calculated for the outer radius of the core in the original model at $10^5$~yr.  The line profiles are plotted in Figure~\ref{fig:cew_temp}, and the ratios of blue-to-red peak intensities for each temperature profile are given in Table~\ref{bluepeaks3}.
 The excitation temperatures of the $3\to2$ transition are plotted in Figure~\ref{Tex}.

From Figure~\ref{fig:cew_temp} we can see that all of the CS transitions now show high velocity wings since CS is not depleted in the  high velocity central core regions. 
For all of the temperature profiles, the $2\to1$ lines are more or less symmetric,  but the $5\to4$ and $3\to2$ lines (which trace the denser, faster moving inner material) show stronger asymmetries.
The red-blue asymmetries in the lines (although not strong) broadly reflect the expected dependency on the  excitation temperature profile.
 The class 0 temperature profile gives the strongest blue asymmetries, which can be explained by the very high excitation temperature in the core centre (where there is still a high abundance of CS since we have neglected depletion).
It is also interesting that {\it even for the constant  kinetic temperature profile}, the $5\to4$ line shows a blue asymmetry.

These results support the interpretation of Zhou et al. (1993), that a  kinetic temperature gradient with a warm core centre  can explain the blue asymmetry for inside-out collapse.  However, they also indicate that, in order for CS to show the blue asymmetry, it cannot be depleted in the core centre.

\begin{table}
\caption{The ratio of blue-to-red peak intensities for different temperature profiles for the inside-out collapse at $1.0 \times 10^5$~yr, assuming a constant CS abundance across the core.}
\begin{center}
\begin{tabular}{l|ccc}
Temp. profile & CS($5\to4$) & CS($3\to2$) & CS($2\to1$) \\ \hline \hline
10~K &1.29 &1.03 & 1.01 \\
Class 0 &1.52 & 1.10&1.03\\
12K centre - 8K edge  &1.29 &1.08 & 1.04\\
8K centre - 12K edge & 1.10&0.95 &0.98\\
\end{tabular}
\end{center}
\label{bluepeaks3}
\end{table}

\section{Discussion}
\label{collapse_disc}

It has been demonstrated in the previous section that the abundance profile of the molecules across the core  (in particular whether or not there is depletion) can have a significant impact upon the line profiles.  We therefore first discuss the sensitivity of the models to different chemical conditions.


\subsection{Initial molecular abundances}
\label{abundances}
We have assumed that, in each of our models, the core begins its collapse from atomic initial conditions (apart from hydrogen which is initially 90\% molecular).  This, however, might not be accurate, especially for the inside-out collapse model which starts from a more evolved state.
This has been recognised in previous studies.
For example, Rawlings et al. (1992) assume that the core is initially at a uniform density of $2.8 \times 10^3$~cm$^{-3}$, with the initial abundances taken from a previous model by \citet{Nejad90},
and then
allow modified free-fall collapse to the isothermal sphere configuration, where the inside-out collapse eventually begins.  In Rawlings \& Yates (2001), they also consider the possibility that once the core has reached the isothermal sphere state, it can remain in pressure-balanced equilibrium prior to the inside-out collapse. 

It is debatable what the initial abundances for such collapse models should be,  since the pre-stellar core lifetime is not yet well determined \citep{Andre09}.
 We have therefore investigated the effect on the line profiles  for both models of keeping the core static for a period of time, $t_\textrm{stat}$, ranging from $10^3$ to $10^8$~yr, allowing the chemistry to reach a more evolved state before the collapse is initiated.
 The drawback of this investigation for the inside-out collapse model is that it may be unrealistic that a core would be static in the  isothermal sphere configuration preceding the onset of the CEW.  Instead, it is more likely that it is contracting  from lower densities, and therefore the chemical timescales could be longer than those presented here.   

\begin{figure}
\begin{center}
\includegraphics[width=90mm]{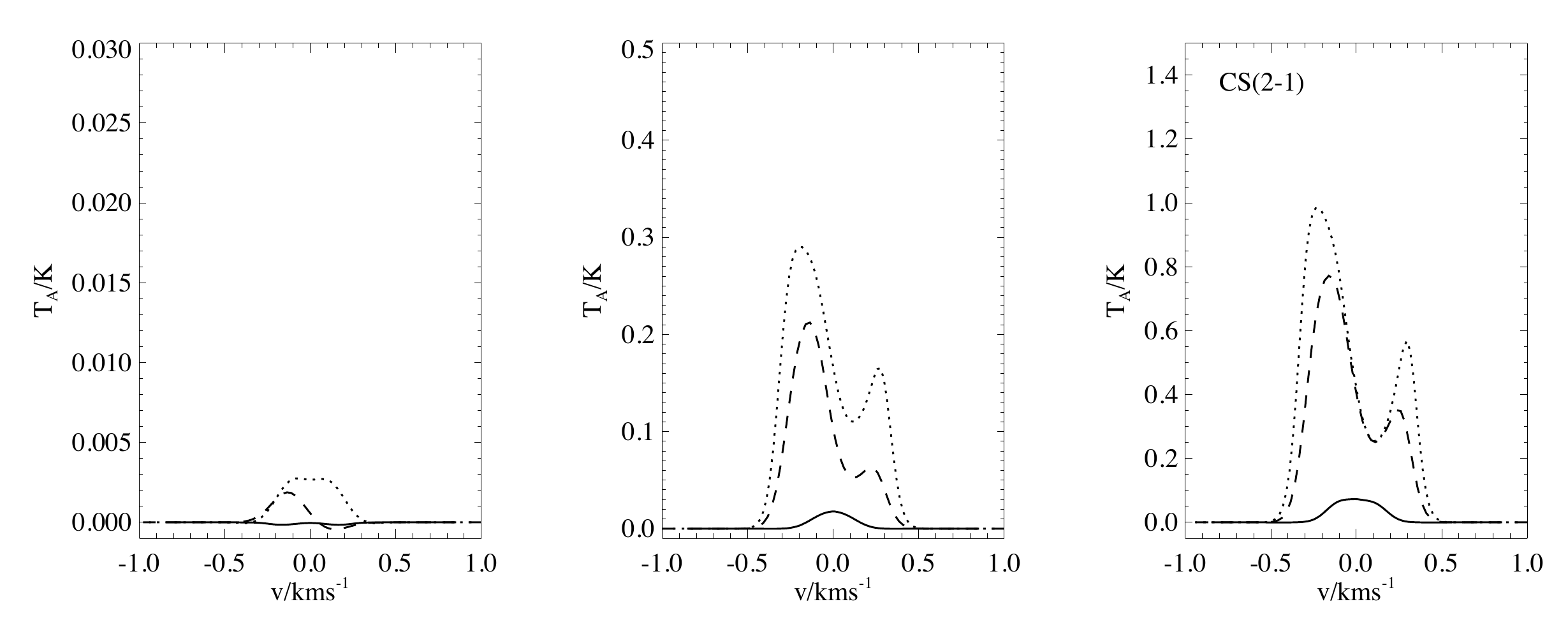}
\caption{Line profiles of CS for transitions $J = 5 \to 4$, $3 \to 2$, and $2 \to 1$ for the ambipolar
diffusion model at $1 \times 10^5$ yr (solid line), $5 \times 10^5$ yr (dotted line) and $1 \times 10^6$ yr (dashed line).  In this model the core has been static for $t_\textrm{stat}=10^5$~yr before it begins collapsing.}
\label{cs_1e5}
\end{center}
\end{figure}

\begin{figure}
\begin{center}
\includegraphics[width=90mm]{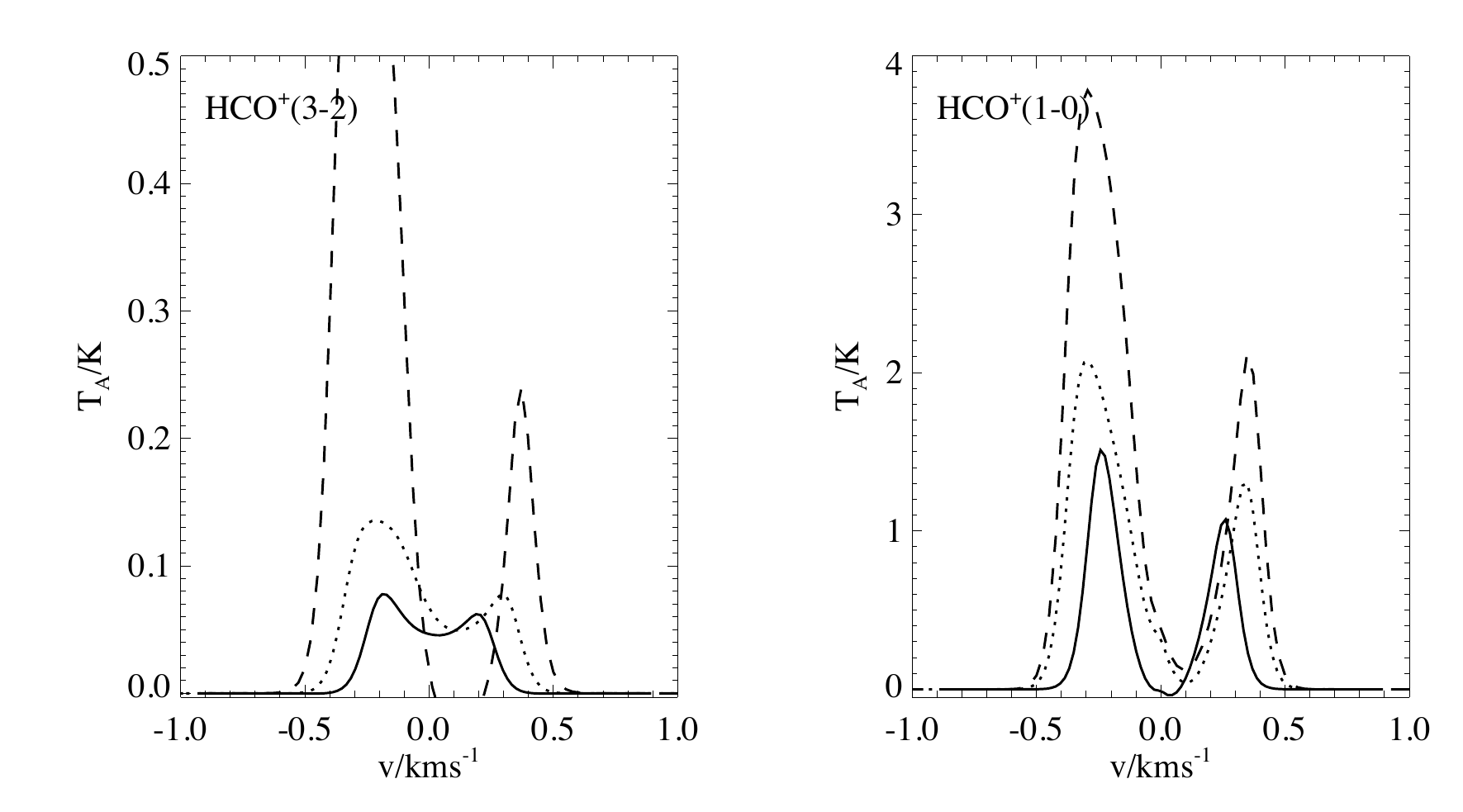}
\caption{Line profiles of HCO$^+$ for transitions $J = 3 \to 2$ and  $1 \to 0$ for the ambipolar
diffusion model at $1 \times 10^5$ yr (solid line), $5 \times 10^5$ yr (dotted line) and $1 \times 10^6$ yr (dashed line).  In this model the core has been static for $t_\textrm{stat}=10^7$~yr before it begins collapsing.}
\label{hcop_1e7}
\end{center}
\end{figure}


For the ambipolar diffusion model, we find that there is no effect on the line profiles presented in Figure~\ref{all_AD} for  $t_\textrm{stat}\lesssim10^4$~yr.
For $t_\textrm{stat}=10^5$~yr, there is an effect on the CS line profiles, shown in Figure~\ref{cs_1e5}. 
In Figure~\ref{all_AD}, the peak intensities of the CS($3\to2$) and ($2\to1$) lines increase as the core evolves from $5\times10^5$~yr to $10^6$~yr, but for $t_\textrm{stat}=10^5$~yr we find that those lines reach their maximum peak intensity at $5\times10^5$~yr, and then they decrease by $10^6$~yr.
This is because for $t_\textrm{stat}=0$, CS reaches its peak abundance between $5.6\times10^5$~yr (for the inner shell) and $7.0\times10^5$~yr (for the outer shell), but for $t_\textrm{stat}=10^5$~yr, the peak abundance is reached at  slightly earlier times of  $5.2\times10^5$ to $6.0\times10^5$~yr.   Therefore for $t_\textrm{stat}=10^5$~yr, by a collapse time of $10^6$~yr, the CS abundance in all shells is already past its peak, resulting in a lower abundance across the core compared to $t_\textrm{stat}=0$.

For $t_\textrm{stat}\geq10^6$~yr,  CS is severely depleted, so 
the peak intensities are very much reduced.
For $t_\textrm{stat}\geq10^7$~yr, CS becomes undetectable.

 For HCO$^+$, the line profiles are not affected until $t_\textrm{stat}=10^6$~yr.  For this $t_\textrm{stat}$, the initial HCO$^+$ abundance for the collapse stage is very high ($3.2\times10^{-9}$ compared to $\sim10^{-20}$ for $t_\textrm{stat}=0$), resulting in much higher peak intensities
for the earlier collapse times ($10^5$ and $5\times10^5$~yr), compared to those with $t_\textrm{stat}=0$.

For $t_\textrm{stat}=10^7$~yr, shown in Figure~\ref{hcop_1e7}, the initial abundance of HCO$^+$ is even higher ($8.7\times10^{-9}$), and line intensities increase even further compared to $t_\textrm{stat}=0$.


By $t_\textrm{stat}=10^8$~yr, however, HCO$^+$ begins to deplete before the collapse begins (the initial HCO$^+$ abundance for the collapse stage is now only $1.3\times10^{-9}$), and the line intensities decrease compared to $t_\textrm{stat}=10^7$~yr.


\begin{figure}
\begin{center}
\includegraphics[width=90mm]{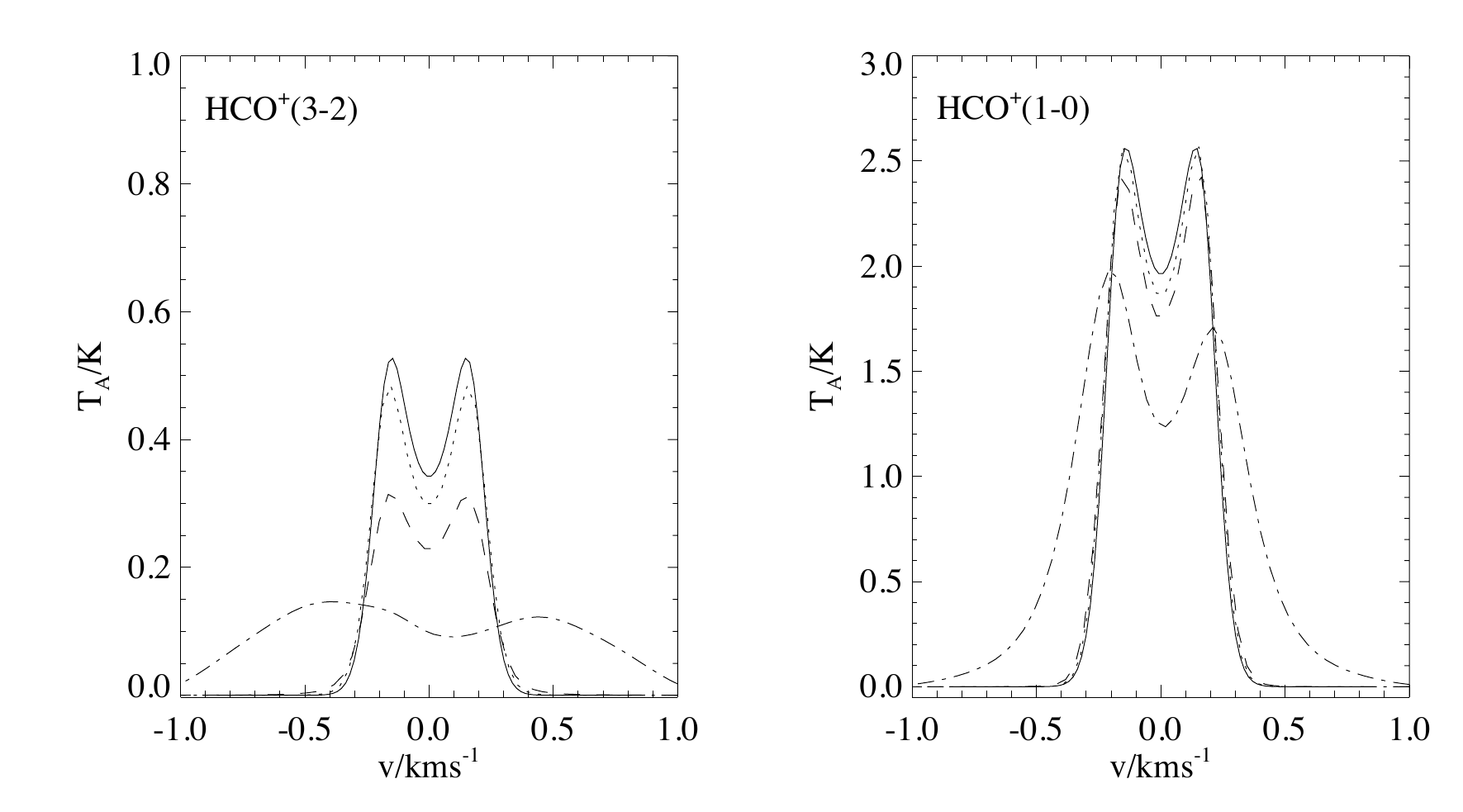}
\caption{Line profiles of HCO$^+$ for transitions $J = 3 \to 2$ and  $1 \to 0$ for the inside-out model at 
 $1 \times 10^4$ yr (solid line), $5 \times 10^4$ yr (dotted line), $1 \times 10^5$ yr (dashed line) and $2.5 \times 10^5$ yr (dot-dashed line).  In this model the core has been static for $t_\textrm{stat}=10^5$~yr before it begins collapsing. }
\label{hcop_cew_1e5}
\end{center}
\end{figure}

 For the inside-out collapse model, there is no effect on the CS or HCO$^+$ line profiles until $t_\textrm{stat}=10^5$~yr.  For this $t_\textrm{stat}$, even for a collapse time of $10^4$~yr, the CS becomes severely depleted in the inner shells (for the three innermost shells the abundances of CS are $4.9\times10^{-15}$, $5.6\times10^{-12}$ and $1.2\times10^{-9}$ for $t_\textrm{stat}=10^5$~yr, compared to $7.6\times10^{-10}$, $1.2\times10^{-8}$ and $1.3\times10^{-8}$ for $t_\textrm{stat}=0$).  This results in large decrease in the line intensity during the collapse for all the CS lines considered.  This effect is particularly severe for the CS($5\to4$) which traces the innermost parts of the shell; for a collapse time of $10^4$~yr the line intensity reduces from 1.2 for $t_\textrm{stat}=0$ to 0.12 for  $t_\textrm{stat}=10^5$~yr.

 For HCO$^+$, for $t_\textrm{stat}=10^5$ (plotted in Figure~\ref{hcop_cew_1e5}), at all times during the collapse the abundances in the outer parts of the core are greater than the abundance in the same shells for $t_\textrm{stat}=0$.  However, at this $t_\textrm{stat}$, the inner shells are more depleted.  Because of this, for $t_\textrm{stat}=10^5$~yr, the HCO$^+$($3\to2$) lines (which trace the inner parts of the core) have lower peak intensities than for $t_\textrm{stat}=0$, and the HCO$^+$($1\to0$) lines (which trace the lower density outer material), show higher peak intensities for $t_\textrm{stat}=10^5$~yr.

By $t_\textrm{stat}=10^6$~yr, the CS lines have practically disappeared because of depletion, and even the HCO$^+$($1\to0$) lines start decreasing in intensity as the HCO$^+$ depletion starts to affect larger radii.

 As with the ambipolar diffusion model, keeping the core static for a period of time affects the CS and HCO$^+$ lines differently.  For both models, including a static period tends to weaken the CS lines due to depletion, but for moderate values of $t_\textrm{stat}$ ($\sim10^6$~yr for the ambipolar diffusion model and $\sim10^5$~yr for the inside-out collapse model), the HCO$^+$ intensities can increase.  This is because, as explained in Section~\ref{chemdisc}, HCO$^+$ is a `depletion enhanced' molecule. 
The CS:HCO$^+$ intensity ratios could therefore be a useful indicator of the initial state of chemical evolution of a collapsing core. The inside-out collapse model results indicate that for CS to be detectable during the collapse phase, the core cannot have been static with the isothermal sphere density configuration for more than $10^6$~yr.

\subsection{Visual extinction and inclusion of photoreactions}

In our models we have assumed that the cores are sufficiently embedded within a larger molecular cloud such that the visual extinction at the edge of the envelope is $> 10$~magnitudes, practically suppressing all of the direct photoreactions.  It is possible, however, that photoreactions play a more significant role in the chemistry, particularly at larger radii where the material is less shielded. This is particularly likely in cases where the parent molecular cloud has been dispersed (eg. the pre-stellar core B68 \citep{Maret07} and the Class 0 object B335 \citep{Zhou93}).

To test the importance of these effects we have re-run our chemical model for the shells at the outer edges of the envelopes for both collapse models. We find that, for $A_\text{v}\leq 5$~mag, the abundances of CS and HCO$^+$ are significantly reduced.  For CS, this is because it is directly destroyed by photons.  
The higher ionisation level suppresses HCO$^+$ through dissociative recombination.

These effects lead to reduced molecular abundances in the outer parts of the core, resulting in steeper gradients in the abundance profiles, and thus
changing the line profile shapes.  For example, in the ambipolar diffusion model, we suggested that the blue asymmetries in CS and HCO$^+$  lines are caused by the redshifted material at the front of the envelope preferentially absorbing the red emission from further inside the cloud. If there was less material in the outer envelope there would be less absorption, and we therefore expect a less pronounced blue asymmetry.  

\subsection{Effect of ionisation fractions for the ambipolar diffusion model}
\label{IF}

In Section~\ref{ADcoll} we mentioned that the ionisation fraction predicted by our chemical model is $\sim4$ times less than the ionisation fraction predicted by the model of SR10 at late times ($10^6$~yr) , for large radii ($>4\times10^{17}$~cm).  Since the dynamics of SR10's model, which we used as an input to our chemical model, depend on the ionisation fraction, it is possible that, to be consistent with our chemical model, the outer parts of the core ($R>4\times10^{17}$~cm) should be less strongly coupled to the magnetic field and could therefore have larger infall velocities.

To understand what effect this would have, we have recalculated the CS line profiles for a collapse time of $10^6$~yr, but doubled the infall velocity for $R>4\times10^{17}$~cm.  We find that even for such high infall velocities in the outer parts of the core, the blue-shifted peak of the CS line profiles is almost completely unaffected.  The red-shifted peaks, on the other hand, show a decrease in peak intensity, and for the CS($2\to1$) line the red-shifted peak is shifted further to the right; in the original spectrum the peak intensity is 0.51~K at 0.26~km~s$^{-1}$, but with the increased velocities in the outer core, the peak intensity decreases to 0.26~K at 0.31~km~s$^{-1}$.

This is the expected behaviour, since the higher velocity red-shifted material at the front of the core (i.e. closest to the observer) is now more strongly radiatively coupled with the red-shifted high velocity material in the core centre, and therefore can absorb more of the red-shifted emission from the core centre.
 Note also that in this `high-velocity' model, the infall velocities are similar to those of \citet{Pav03} ($\sim 0.2$~km~s$^{-1}$), and the blue-to-red peak intensity ratios for the CS($2\to1$) line predicted by both models  are now in very good agreement with each other ($\sim4$).

\section{Conclusions}
\label{conc}

In this paper we have modelled and compared the line profiles of several transitions of CS and HCO$^+$ at various times for cores undergoing (a) ambipolar diffusion regulated collapse and (b) inside-out collapse. Rather than assuming a constant abundance for each molecule, we have coupled a chemical model to the dynamical models to try to generate realistic, time dependent abundance profiles, which are then used as an input to the radiative transfer code to calculate the line profiles.  
We have found four important conclusions:

\begin{enumerate}

\item  For the inside-out collapse model to exhibit the blue asymmetry it is necessary to suppress freeze-out {\it and} impose a negative temperature gradient on the core, whereas the ambipolar diffusion model exhibits the blue asymmetry even when there is significant freeze-out, and for all the kinetic temperature profiles tested.
%
%
%
This conclusion is important because, for example, CS has been observed to be significantly depleted within dense cores (e.g. Bergin et al. 2001; Pagani et al. 2005), and yet exhibits the blue asymmetric line profile (e.g. Mardones et al., 1997; Lee et al., 1999).  Zhou et al (1993) found that their observations of CS in B335 were very well fitted by the inside-out collapse model, but in these studies the CS abundance was assumed to be constant across the core.  Given the strong evidence for depletion of CS in such cores, it may be worth investigating in more detail the CS abundance profile in this core using interferometric observations, and seeing if the inside-out model is still able to reproduce the CS observations with a realistic abundance profile.
 Note that Belloche et al. (2002) also found that the inside-out collapse model could not reproduce the blue asymmetry  observed towards the class 0 protostar IRAM 04191 unless the collapse expansion wave radius is very large ($\ge$ 10 000 AU), but then this resulted in very high velocities at small radii, giving rise to excessively large linewidths.  They estimated that CS is depleted by a factor of $\sim20$ in the centre of this core.  Since they took into account this depletion in their radiative transfer model, this could explain why their inside-out collapse model failed to reproduce the observations.
%
%

The fact that a negative kinetic temperature gradient is required in the inside-out collapse model but not in the ambipolar diffusion model is due to the different line formation processes:
%
%
in the ambipolar diffusion model the blue asymmetry arises because the foreground absorbing material is also infalling and redshifted, but for the inside-out collapse model it is because the blue emission originates from the central regions of the core where the excitation temperature is relatively high, whereas the red emission originates from further away from the centre where there is relatively lower excitation temperature.

\item There are marked differences between the two models in the line strengths, asymmetries and time-dependencies of the line profiles. For the ambipolar diffusion model, the CS($2\to1$) and ($3\to2$) line profiles are strong and demonstrate a blue asymmetry (with an absorption trough that is not at zero velocity). The CS($5\to4$) line is not detectable. The HCO$^+$($3\to2$) line is only detectable after $\sim 10^6$ years, whilst the ($1\to0$) line strength strongly increases with time. 
For the inside-out model, all three CS lines are strong from the earliest stages, but decline with time. The HCO$^+$($1\to0$) line is also strong at all times, but the strength of the ($3\to2$) line steadily declines. 
Thus, subject to caveats about possible degeneracies, these various characteristics may be useful in determining the dynamical structure and evolutionary status of low-mass infall regions.

\item The initial molecular abundances in collapsing cores affect the evolution of the line profiles. 
 In particular, if the core is held static for a period of time before the collapse, the peak intensity ratio between the CS and HCO$^+$ lines can be strongly affected.
This is a major source of uncertainty in the theoretical interpretation of observed line profiles.

\end{enumerate}

This work demonstrates that the inside-out collapse model of \citet{Shu77}, often referred to as the `standard' infall model \citep[e.g.][]{Dunham10,Hung10}, has difficulty reproducing the blue asymmetry in the line profiles when the chemical evolution of the core is taken into account, due to depletion in the core centre and a lack of extended inward motions.  The ambipolar diffusion model does not suffer from this problem, because  even when the core has reached an advanced, centrally condensed state, there are still extended inward motions.

 Before using line profiles as diagnostics of a particular collapse model, it is essential to investigate the abundance profile of the tracer molecule, in particular to see if there is any evidence for depletion.

\section{Acknowledgements}J.F.R. acknowledges the support of the MICINN under grant number ESP2007-65812-C02-C01.  We would like to thank the referee for suggestions which helped to improve the final version of this paper significantly.

\label{lastpage}


\begin{thebibliography}{}
\bibitem[\protect\citeauthoryear{Aikawa et al.}{2003}]{Aikawa03}Aikawa, Y.,  Ohashi, N.,  Herbst, E., 2003, ApJ, 593, 906
\bibitem[\protect\citeauthoryear{Andr\'e et al.}{2009}]{Andre09}Andr{\'e}, P., Basu, S. \& Inutsuka, S., 2009, 
The Formation and Evolution of Prestellar Cores, Structure Formation in Astrophysics, Cambridge University Press
\bibitem[\protect\citeauthoryear{Belloche et al.}{2002}]{Belloche02}Belloche, A., Andr\'e, P., Despois, D., \& Blinder, S., 2002, A\&A, 393, 927
\bibitem[\protect\citeauthoryear{Bergin et al.}{2001}]{Bergin01}Bergin, E. A., Ciardi, D. R., Lada, C. J., Alves, J., Lada, E. A., 2001, ApJ, 557, 209
\bibitem[\protect\citeauthoryear{Buch \& Zhang}{1991}]{Buch91}Buch, V., Zhang, Q., 1991, ApJ, 379, 647
\bibitem[\protect\citeauthoryear{Caselli et al.}{2002}]{Caselli02} Caselli, P., Walmsley, C. M., Zucconi, A., Tafalla, M., Dore, L., \& Myers, P. C., 2002, ApJ, 565, 344
\bibitem[\protect\citeauthoryear{Crutcher et al.}{1999}]{Crutcher99} Crutcher, R. M., 1999, ApJ, 520, 706
\bibitem[\protect\citeauthoryear{Cuppen et al.}{2009}]{Cuppen09}Cuppen, H. M., van Dishoeck, E. F., Herbst, E., Tielens, A. G. G. M., 2009, A\&A, 509, 275
\bibitem[\protect\citeauthoryear{Duley \& Williams}{1984}]{Duley84}Duley, W. W., \& Williams, D. A., 1984, Interstellar Chemistry, Academic Press, London
\bibitem[\protect\citeauthoryear{Dunham et al.}{2010}]{Dunham10}Dunham, M. M., Evans, N. J., Terebey, S., Dullemond, C. P., Young, C. H., 2010, ApJ, 710, 470
\bibitem[\protect\citeauthoryear{Evans et al.}{2005}]{Evans05}Evans, II, N. J., Lee, J.-E., Rawlings, J. M. C., Choi, M., 2005, ApJ, 626, 919
\bibitem[\protect\citeauthoryear{Evans et al.}{1994}]{Evans94}Evans, II, N. J., Zhou, S., Koempe, C. \& Walmsley, C. M., 1994, ApSS, 212, 139
\bibitem[\protect\citeauthoryear{Fuchs et al.}{2009}]{Fuchs09}Fuchs, G. W., Cuppen, H. M., Ioppolo, S., Romanzin, C., Bisschop, S. E., Andersson, S., van Dishoeck, E. F., Linnartz, H., 2009, A\&A, 505, 629
\bibitem[\protect\citeauthoryear{Garrod et al.}{2008}]{Garrod08} Garrod, R. T., Widicus Weaver, S. L., \&Herbst, E., 2008, ApJ, 682, 283
\bibitem[\protect\citeauthoryear{Hung et al.}{2010}]{Hung10}Hung, C. L., Lai, S. P.,  Yan, C. H., 2010, ApJ, 710, 207
\bibitem[\protect\citeauthoryear{Hunter}{1977}]{Hunter77} Hunter, C., 1977, ApJ, 218, 834
\bibitem[\protect\citeauthoryear{Larson}{1969}]{Larson69}Larson, R. B., 1969, MNRAS, 145, 271
\bibitem[\protect\citeauthoryear{Lee et al.}{2004}]{Lee04}Lee, C. W., Myers, P. C., Plume, R., 2004, JKAS, 37, 257
\bibitem[\protect\citeauthoryear{Lee et al.}{2001}]{Lee01}Lee, C. W., Myers, P. C., Tafalla, M., 2001, ApJS, 136, 703
\bibitem[\protect\citeauthoryear{Lee et al.}{1999}]{Lee}Lee, C. W., Myers, P. C., Tafalla, M., 1999, ApJ, 526, 788
\bibitem[\protect\citeauthoryear{L'eger et al.}{1985}]{Leger85}L\'eger A., Jura M., Omont A., 1985, A\&A, 144, 147 
\bibitem[\protect\citeauthoryear{Leung \& Brown}{1977}]{Leung97}Leung, C. M., Brown, R. L., 1977, ApJ, 214, 73
\bibitem[\protect\citeauthoryear{Mardones et al.}{1997}]{Mardones}Mardones, D., Myers, P. C., Tafalla, M., Wilner, D. J., 1997, ApJ, 489, 719
\bibitem[\protect\citeauthoryear{Maret \& Bergin}{2007}]{Maret07} Maret, S., \& Bergin, 
E. A., 2007, ApJ, 664, 956
\bibitem[\protect\citeauthoryear{Mathis et al.}{1983}]{Mathis83}Mathis, J. S.,  Mezger, P. G., \& Panagia, N., 1983, A\&A, 128, 21	
\bibitem[\protect\citeauthoryear{Millar et al.}{1997}]{Millar07}Millar, T. J., Farquhar, P. R. A., Willacy, K., 1997, A\&AS, 121, 139
\bibitem[\protect\citeauthoryear{Motte \& Andr'e}{2001}]{Motte}Motte, F., Andr{\'e}, P., 2001, A\&A, 365, 440 
\bibitem[\protect\citeauthoryear{Mouschovias \& Spitzer}{1976}]{Mousc76}Mouschovias, T. C. \& Spitzer, Jr., L., 1976, ApJ, 210, 326 
\bibitem[\protect\citeauthoryear{Nejad et al.}{1990}]{Nejad90}Nejad, L. A. M.,  Williams, D. A., \& Charnley, S. B., 1990, MNRAS, 246, 183
\bibitem[\protect\citeauthoryear{Pagani et al.}{2005}]{Pagani}Pagani, L., Pardo, J.-R., Apponi, A. J-. Bacmann, A., Cabrit, S., 2005, A\&A, 429, 181
\bibitem[\protect\citeauthoryear{Pavlyuchenkov et al.}{2003}]{Pav03}Pavlyuchenkov, Y. N., Shustov, B. M., Shematovich, V. I., Wiebe, D. S., Li, Z-Y., 2003, ARep, 47, 176 
\bibitem[\protect\citeauthoryear{Penston}{1969}]{Penston69}Penston, M. V., 1969, MNRAS, 144, 425
\bibitem[\protect\citeauthoryear{Rawlings et al.}{1992}]{Rawlings92}Rawlings, J. M. C., Hartquist, T. W., Menten, K. M. \& Williams, D. A., 1992, MNRAS, 255, 471
\bibitem[\protect\citeauthoryear{Rawlings et al.}{2001}]{Rawlings01}Rawlings, J. M. C.,  Yates, J., 2001, MNRAS, 326, 1423
\bibitem[\protect\citeauthoryear{Roberts et al.}{2007}]{Roberts07}Roberts, J. F., Rawlings, J. M. C., Viti, S., Williams, D. A., 2007, MNRAS, 382, 733
\bibitem[\protect\citeauthoryear{Shematovich et al.}{2003}]{Shem03}Shematovich, V. I., Wiebe, D. S., Shustov, B. M., Li, Zhi-Yun, 2003, ApJ, 588, 894
\bibitem[\protect\citeauthoryear{Shu}{1977}]{Shu77}Shu, F. H., 1977, ApJ, 214, 488
\bibitem[\protect\citeauthoryear{Tafalla et al.}{2004}]{Tafalla04}Tafalla, M., Myers, P. C., Caselli, P.,  Walmsley, C. M., 2004, A\&A, 416, 191  
\bibitem[\protect\citeauthoryear{Tsamis et al.}{2008}]{Tsamis08}Tsamis, Y. G., Rawlings, J. M. C., Yates, J. A., Viti, S., 2008, MNRAS, 388, 898
\bibitem[\protect\citeauthoryear{van Zadelhoff et al.}{2002}]{vZad02}van Zadelhoff, G.-J., Dullemond, C. P., van der Tak, F. F. S., Yates, J. A., Doty, S. D., Ossenkopf, V., Hogerheijde, M. R., Juvela, M., Wiesemeyer, H.,  Sch{\"o}ier, F. L., 2002, A\&A, 395, 373 
\bibitem[\protect\citeauthoryear{Ward-Thompson et al.}{1994}]{WT94}Ward-Thompson, D., Scott, P. F., Hills, R. E., Andr'e, P., 1994, MNRAS, 268, 276
\bibitem[\protect\citeauthoryear{Weidenschilling \& Ruzmaikina}{1994}]{Weid94}Weidenschilling, S. J. \& Ruzmaikina, T. V., 1994, ApJ, 430, 713 
\bibitem[\protect\citeauthoryear{Whitworth \& Summers}{1985}]{Whitworth85}Whitworth, A., Summers, D., 1985, MNRAS, 214, 1
\bibitem[\protect\citeauthoryear{Woodall et al.}{2007}]{Woodall07}Woodall, J., Ag{\'u}ndez, M., Marwick-Kemper, A. J., Millar, T. J., 2007, A\&A, 466, 1197 
\bibitem[\protect\citeauthoryear{Zhou et al.}{1990}]{Zhou90}Zhou, S., Evans, II, N. J., Butner, H. M., Kutner, M. L., Leuing, C. M., Mundy, L. G., 1990., ApJ, 363, 168
\bibitem[\protect\citeauthoryear{Zhou et al.}{1993}]{Zhou93}Zhou, S., Evans, II, N. J., Koempe, C.,  Walmsley, C. M., 1993, ApJ, 404, 232




\end{thebibliography}
\end{document}